\documentclass[%
 reprint,
nofootinbib,
amsmath,amssymb,citeautoscript,
aip,
floatfix,
]{revtex4-2}
\pdfoutput=1
\usepackage{titlesec}
\titleformat*{\section}{\normalsize\bfseries}
\titleformat*{\subsection}{\normalsize\itshape}
\titlespacing*{\section} {0pt}{3ex}{1ex}
\titlespacing*{\subsection} {0pt}{2ex}{0ex}
\usepackage{caption}
\usepackage{graphicx}
\usepackage{dcolumn}
\usepackage{natbib}
\usepackage{bm}
\usepackage{amsbsy}

\DeclareUnicodeCharacter{2061}{}
\usepackage{chemist}
\usepackage[version=4]{mhchem}
\makeatletter
\newcommand{\fmarki}{*}
\newcommand{\fmarkii}{\ensuremath{\dagger}}
\newcommand{\fmarkiii}{\ensuremath{\ddagger}}
\def\@alph#1{{\ifcase#1\or \fmarki\or \fmarkii\or \fmarkiii\or \fmarkiv\or \fmarkv\or \fmarkvi\or \fmarkvii\or \fmarkviii\or \fmarkix \else\@ctrerr\fi}}
\makeatother

\newcommand{\smallsim}{\smallsym{\mathrel}{\sim}}

\makeatletter
\newcommand{\smallsym}[2]{#1{\mathpalette\make@small@sym{#2}}}
\newcommand{\make@small@sym}[2]{%
  \vcenter{\hbox{$\m@th\downgrade@style#1#2$}}%
}
\newcommand{\downgrade@style}[1]{%
  \ifx#1\displaystyle\scriptstyle\else
    \ifx#1\textstyle\scriptstyle\else
      \scriptscriptstyle
  \fi\fi
}
\makeatother

\renewcommand{\fmarki}{\textdagger}
\renewcommand{\fmarkii}{a}
\renewcommand{\fmarkiii}{b}
\begin{document}
\title{Atomic evolution of hydrogen intercalation wave dynamics
in palladium nanocrystals}

\author{Daewon Lee}
 \thanks{These authors contributed equally to this work.}
 \affiliation{\mbox{Materials Sciences Division, Lawrence Berkeley National Laboratory, Berkeley, CA 94720, USA}}
 \affiliation{\mbox{Department of Materials Science and Engineering, University of California, Berkeley, CA 94720, USA}}
\author{Sam Oaks-Leaf}
 \thanks{These authors contributed equally to this work.}
 \affiliation{\mbox{Department of Chemistry, University of California, Berkeley, CA 94720, USA}}
\author{Sophia B. Betzler}
 \thanks{These authors contributed equally to this work.}
 \affiliation{\mbox{Materials Sciences Division, Lawrence Berkeley National Laboratory, Berkeley, CA 94720, USA}}
\author{Yifeng Shi}
 \affiliation{\mbox{School of Chemical and Biomolecular Engineering, Georgia Institute of Technology, Atlanta, GA 30332, USA}}
 \author{Siyu Zhou}
 \affiliation{\mbox{School of Chemical and Biomolecular Engineering, Georgia Institute of Technology, Atlanta, GA 30332, USA}}
\author{Colin Ophus}
 \affiliation{\mbox{Department of Materials Science and Engineering, Stanford University, Stanford, CA 94305, USA}}
\author{Lin-Wang Wang}
 \affiliation{\mbox{Key Laboratory of Optoelectronic Materials and Devices, Chinese Academy of Science, Beijing 100083, China}}
\author{Mark Asta}
\affiliation{\mbox{Materials Sciences Division, Lawrence Berkeley National Laboratory, Berkeley, CA 94720, USA}}
\affiliation{\mbox{Department of Materials Science and Engineering, University of California, Berkeley, CA 94720, USA}}
\author{Younan Xia}
 \affiliation{\mbox{School of Chemical and Biomolecular Engineering, Georgia Institute of Technology, Atlanta, GA 30332, USA}}
 \affiliation{The Wallace H. Coulter Department of Biomedical Engineering, Georgia Institute of Technology and Emory University, Atlanta, GA 30332, USA}
\affiliation{\mbox{School of Chemistry and Biochemistry, Georgia Institute of Technology, Atlanta, GA 30332, USA}}
\author{David T. Limmer\textsuperscript{*}}
\email{dlimmer@berkeley.edu }
\affiliation{\mbox{Materials Sciences Division, Lawrence Berkeley National Laboratory, Berkeley, CA 94720, USA}}
\affiliation{\mbox{Department of Chemistry, University of California, Berkeley, CA 94720, USA}}
\affiliation{\mbox{Chemical Sciences Division, Lawrence Berkeley National Laboratory, Berkeley, CA 94720, USA}}
\affiliation{\mbox{Kavli Energy Nanoscience Institute, Berkeley, CA 94720, USA}}
\author{Haimei Zheng\textsuperscript{*}}
\email{hmzheng@lbl.gov }
\affiliation{\mbox{Materials Sciences Division, Lawrence Berkeley National Laboratory, Berkeley, CA 94720, USA}}
\affiliation{\mbox{Department of Materials Science and Engineering, University of California, Berkeley, CA 94720, USA}}

\date{\today}

\begin{abstract}
Solute-intercalation-induced phase separation creates spatial heterogeneities in host materials, a phenomenon ubiquitous in batteries, hydrogen storage, and other energy devices. Despite many efforts, probing intercalation processes at the atomic scale has been a significant challenge. We study hydrogen (de)intercalation in palladium nanocrystals as a model system and achieve atomic-resolution imaging of hydrogen intercalation wave dynamics by utilizing liquid-phase transmission electron microscopy. Our observations reveal that intercalation wave mechanisms, instead of shrinking-core mechanisms, prevail at ambient temperature for palladium nanocubes ranging from $\smallsim$60 nm down to $\smallsim$10 nm. We uncover the atomic evolution of hydrogen intercalation wave transitioning from non-planar and inclined boundaries to those closely aligned with \{100\} planes. Our kinetic Monte Carlo simulations demonstrate the observed intercalation wave dynamics correspond to sorption pathways minimizing the lattice mismatch strain at the phase boundary. Unveiling the atomic intercalation pathways holds profound implications for engineering intercalation-mediated devices and advancements in energy sciences.
\end{abstract}
\maketitle


\section*{\label{sec:main}Main}

Solute intercalation into host materials is fundamental to energy storage technologies, including lithium-ion batteries\cite{dreyer_thermodynamic_2010, sood_electrochemical_2021} and hydrogen storage\cite{pundt_hydrogen_2004, griessen_thermodynamics_2016}. The intercalated solute species in critical energy storage materials, such as lithium iron phosphate ($\mathrm{Li}_x\mathrm{FePO}_4$, LFP)\cite{lim_origin_2016, li_electrochemical_2018,koo_dynamic_2023}, lithium cobalt oxide ($\mathrm{Li}_x\mathrm{CoO}_2$, LCO)\cite{fraggedakis_scaling_2020,merryweather_operando_2021}, as well as palladium hydride ($\mathrm{PdH}_x$)\cite{li_electrochemical_2018, li_hydrogen_2014, baldi_situ_2014}, induce phase separation into phases with different stoichiometries. This phase separation brings about spatial heterogeneities during solute-induced phase transformations. Both theoretical and experimental studies have established two representative phase transformation mechanisms linked to distinct phase morphologies\cite{koo_dynamic_2023, fraggedakis_scaling_2020, merryweather_operando_2021, singh_intercalation_2008, ulvestad_three-dimensional_2017, park_fictitious_2021}. The diffusion-limited shrinking-core mechanism exhibits a phase boundary propagating inward from all reactive surfaces. In contrast, the insertion-limited intercalation wave mechanism is characterized by one or a few phase boundaries that nucleate and propagate across the particle. These phase boundary morphologies profoundly affect the performance and durability of energy devices\cite{fraggedakis_scaling_2020, ulvestad_three-dimensional_2017, tang_electrochemically_2010, nadkarni_modeling_2019}. Therefore, direct observation of spatial heterogeneities during the phase transformations is crucial for discerning the underlying intercalation principles and guiding the design and engineering of energy storage systems.

Hydrogen intercalation in Pd serves as an archetypal process to understand solute-intercalation-induced phase transformations. Two immiscible phases form when hydrogen is absorbed into a Pd lattice: dilute $\alpha$-$\mathrm{PdH}_x$ and hydrogen-rich $\beta$-$\mathrm{PdH}_x$. Hydrogen-induced phase transformations between these phases have exhibited size-dependent thermodynamics and kinetics\cite{griessen_thermodynamics_2016, baldi_situ_2014,yamauchi_nanosize_2008, ingham_particle_2008, bardhan_uncovering_2013, syrenova_hydride_2015}. Ensemble-level studies on Pd nanoparticles with sub-10 nm diameter have suggested the shrinking-core mechanism to explain the observed (de)hydriding behaviors\cite{langhammer_size-dependent_2010, langhammer_kinetics_2010}. However, in situ single-particle imaging techniques have directly visualized the intercalation wave mechanism (or spherical-cap phase geometry) during hydrogen absorption and desorption in Pd nanocubes or nanoparticles $\smallsim$20 nm or larger\cite{ulvestad_three-dimensional_2017, ulvestad_avalanching_2015, narayan_direct_2017, sytwu_visualizing_2018, vadai_-situ_2018}. These pioneering studies indicate an unexplored size regime, where for palladium nanocrystals smaller than $\smallsim$20 nm, it is unclear whether hydrogen-induced phase transformations follow the shrinking-core or intercalation wave mechanism. Moreover, despite extensive prior efforts\cite{griessen_thermodynamics_2016, li_hydrogen_2014, baldi_situ_2014, ulvestad_three-dimensional_2017, ingham_particle_2008, bardhan_uncovering_2013, syrenova_hydride_2015, ulvestad_avalanching_2015, narayan_direct_2017, sytwu_visualizing_2018, vadai_-situ_2018, liu_nanoantenna-enhanced_2011, johnson_facets_2019, koo_ultrathin_2024} to study hydrogen sorption behaviors in Pd nanoparticles, in situ atomic-resolution imaging of hydrogen-induced phase transformations in Pd nanocrystals has not been achieved due to technical challenges.

Herein, we utilize liquid-phase transmission electron microscopy (TEM) to directly track hydrogen (de)intercalation processes in single Pd nanocrystals at atomic resolution. The electron-beam-induced radiolysis of the aqueous solution generates hydrogen gas within a liquid cell, and the varying amounts of generated hydrogen control whether the Pd nanocrystals undergo hydrogen absorption or desorption. We visualize hydrogen sorption processes in individual Pd nanocubes ranging from $\smallsim$10 to $\smallsim$60 nm, with a primary focus on the uncharted size range down to $\smallsim$10 nm. Our observations confirm that hydrogen-induced phase transformations in all tested nanocrystals follow the intercalation wave mechanism. Furthermore, this unprecedented imaging capability allows us to unveil the structural dynamics of hydrogen intercalation waves at the atomic level. Using kinetic Monte Carlo simulations of a simple elastic Ising model Hamiltonian, we are able to recapitulate the basic experimental observations. This enables us to ascribe the experimentally observed intercalation wave dynamics to the minimization of elastic energies, which holds down to nanocubes smaller than 10 nm. 

\section*{\label{sec:results}Results and discussion}
\subsection*{\label{sec:ltem}Liquid-phase TEM for investigating hydrogen sorption behaviors}
\textbf{Fig. 1a} illustrates our experimental setup using the liquid-phase TEM to study hydrogen sorption behaviors in single Pd nanocrystals. We synthesize \{100\}-enclosed Pd nanocubes with sizes ranging from $\smallsim$10 to $\smallsim$60 nm, following established synthetic \cite{sytwu_visualizing_2018, jin_synthesis_2011} (see \textbf{Methods}). \textbf{Figs. 1b and 1c} display a high-resolution TEM (HRTEM) image with its corresponding fast Fourier transform (FFT) pattern along the [00-1] zone axis, as well as a scanning TEM (STEM) image and its respective Pd-L lines energy-dispersive X-ray spectroscopy (EDS) map for the representative Pd nanocubes in the $\smallsim$10 to $\smallsim$30 nm size range. Similar S/TEM characterization results for Pd nanocubes in the bigger size range are provided in \textbf{Supplementary Fig. 1}.
To prepare samples for liquid-phase TEM experiments, the as-prepared Pd nanocubes are dispersed on TEM grids coated with electron-transparent membranes (see \textbf{Methods} and \textbf{Supplementary Fig. 2} ). A droplet of aqueous KOH solution is then loaded into a liquid cell made of two sandwiched TEM grids (one with the dispersed Pd nanocubes), encapsulating a thin liquid film inside the cell\cite{yin_visualization_2019, zhang_defect-mediated_2022}, (see \textbf{Methods}). During liquid-phase TEM imaging, the radiolysis of water produces various radiolytic products, with hydrogen and oxygen gases being the dominant ones\cite{hart_hydrated_1964, grogan_bubble_2014, schneider_electronwater_2014}. By adjusting the electron dose rate, we can control the level of hydrogen generation inside the liquid cell (see \textbf{Supplementary Note 1}  and \textbf{Supplementary Figs. 3 and 4}). The differing hydrogen amount dictates the hydrogen (de)intercalation processes in Pd nanocrystals.

First, we examine $\smallsim$20 to $\smallsim$60-nm Pd nanocubes to validate that our liquid-phase TEM setup is well-suited for studying hydrogen-induced phase transformations. \textbf{Extended Data} \textbf{Fig. 1a} shows a TEM image of a Pd nanocube captured during the initial stage of hydrogen absorption, which features dark fringes at the upper right corner. The phase transformation from $\alpha$-$\mathrm{PdH}_x$ to $\beta$-$\mathrm{PdH}_x$ is accompanied by a $\smallsim$3.5\% increase in the lattice constant\cite{wyckoff_structure_1924, wicke_hydrogen_1978}, causing inward shifts in the radial distances of reflections in the FFT patterns (described in \textbf{Supplementary Fig. 5} ). The respective FFT pattern exhibits the coexistence of $\alpha$- and $\beta$-$\mathrm{PdH}_x$ phases with two (-220) spots corresponding to those phases.

Our image analysis, based on geometric phase analysis\cite{hytch_quantitative_1998} (GPA), tracks the spatial distributions of the $\alpha$- and $\beta$-$\mathrm{PdH}_x$ phases within Pd nanocubes during phase transformations (see \textbf{Methods}). The \textit{d}-spacing colormap (\textbf{Extended Data} \textbf{Fig. 1b}) demonstrates that the localized lattice expansion ($\beta$-$\mathrm{PdH}_x$ phase) coincides with the dark fringes in the image. Similar fringes were previously observed during hydrogen absorption using gas environmental STEM\cite{narayan_direct_2017}. These fringes are attributed to the strain caused by the lattice mismatch between $\alpha$- and $\beta$-$\mathrm{PdH}_x$ and/or thickness variations due to the local swelling of the nanocube.
We utilize these fringes to identify the mechanisms of hydrogen-induced phase transformations at lower magnifications. \textbf{Extended Data} \textbf{Fig. 1c} and \textbf{Extended Data Fig. 2} , based on \textbf{Supplementary Videos 1-3}, present the hydrogen intercalation wave mechanisms during hydrogen absorption (illustrated in the left scheme of \textbf{Fig. 1d}) in single Pd nanocubes ranging from $\smallsim$20 to $\smallsim$60 nm. The $\beta$-$\mathrm{PdH}_x$ phase nucleates at one of the corners of the Pd nanocubes and expands until it reaches one of the \{100\} facets, forming $\alpha$/$\beta$-$\mathrm{PdH}_x$ phase boundaries across the nanocubes. These phase boundaries then propagate along one of the $\langle 100\rangle$ directions to complete the hydriding process (also see \textbf{Supplementary Note 2}  for detailed propagation dynamics related to \textbf{Extended Data} \textbf{Fig. 1c}). The observed hydrogen intercalation wave mechanisms are consistent with preceding studies of hydrogen absorption in Pd nanocubes $\smallsim$20 nm or larger in a gas atmosphere\cite{narayan_direct_2017,sytwu_visualizing_2018}, confirming that liquid-phase TEM is a unique approach for revealing hydrogen-induced phase transformations.

\subsection*{\label{sec:direct}Direct visualization of intercalation waves in uncharted Pd size ranges}
Direct atomic-resolution visualization of hydrogen-induced phase transformation dynamics in Pd nanocrystals has been exceptionally challenging due to technical difficulties, including electron-beam-enhanced fugacity of $\mathrm{H}_2$ gas in TEM, which accelerates the hydrogen absorption rate\cite{bond_determination_1986, field_-situ_1997, xie_situ_2015}, and size-dependent hydrogen loading behaviors in Pd nanocubes\cite{griessen_thermodynamics_2016, baldi_situ_2014, bardhan_uncovering_2013, syrenova_hydride_2015}. By fine-tuning the experimental conditions (see \textbf{Methods}), our liquid-phase TEM platform enables atomic-resolution HRTEM imaging of the (de)hydriding processes in Pd nanocubes, with a particular focus on the uncharted $\smallsim$10 to $\smallsim$20 nm size regime (depicted in \textbf{Fig. 2a}). The raw HRTEM images acquired during hydrogen absorption and desorption exhibit excellent atomic resolution. To further enhance the signal-to-noise ratio, we average two successive HRTEM images and perform non-linear Fourier filtering of each averaged image to attenuate its Gaussian noise (see \textbf{Methods} and \textbf{Supplementary Fig. 6}).
Sequential HRTEM images and corresponding \textit{d}-spacing maps reveal spatial distributions of $\alpha$- and $\beta$-$\mathrm{PdH}_x$ phases in 11-24-nm Pd nanocubes during hydrogen (de)intercalation (\textbf{Figs. 2b-2e}, based on \textbf{Supplementary Videos 4-7}). The \textit{d}-spacing colormaps for 22 × 24 nm and 19 × 21 nm nanocubes (\textbf{Figs. 2b and 2c} and \textbf{Extended Data Fig. 3}) confirm that hydrogen absorption proceeds via hydrogen intercalation wave propagation, which aligns with our lower magnification observation in the 20 × 20 nm nanocube (\textbf{Extended Data Fig. 2}).

In \textbf{Figs. 2d and 2e}, we present the direct visualization of hydrogen (de)intercalation processes in Pd nanocubes within the unexplored size range. Interestingly, we observe the intercalation wave mechanisms, rather than the shrinking-core structures, during (de)hydriding. Hydrogen absorption in a 14 × 16 nm nanocube (\textbf{Fig. 2d}) is characterized by the propagation of a hydrogen intercalation wave (i.e., $\alpha$/$\beta$-$\mathrm{PdH}_x$ phase boundary) along the [010] direction. Furthermore, hydrogen desorption in an 11 × 13 nm Pd nanocube (\textbf{Fig. 2e}) features the $\alpha$/$\beta$-$\mathrm{PdH}_x$ interface spanning the entire nanocube. This interface subsequently propagates in both the [100] and [0-10] directions, eventually leading to corner-localized $\beta$-$\mathrm{PdH}_x$, the reverse counterpart of the observed absorption process. These results corroborate that the intercalation wave mechanism persists across all examined size ranges, even down to $\smallsim$10 nm, expanding our understanding of hydrogen-induced phase transformations into the unexplored size regime.

\subsection*{\label{sec:atomic}Atomic evolution of hydrogen intercalation wave dynamics}
Beyond revealing the phase morphologies, our advanced imaging capability allows us to track the atomic-level structural dynamics of hydrogen intercalation waves (schematic on the right in \textbf{Fig. 2a}) as they traverse along the $\langle 100\rangle$ directions. \textbf{Fig. 3a}  presents representative atomic-resolution HRTEM images of the 14 × 16 nm Pd nanocube, capturing the [010] propagation of the hydrogen intercalation wave during absorption. Corresponding Bragg-filtered lattice points (\textbf{Fig. 3b}  and \textbf{Supplementary Fig. 7}) and \textit{d}-spacing colormaps (\textbf{Fig. 3c} ) highlight the $\alpha$/$\beta$-$\mathrm{PdH}_x$ interface regions within the nanocube. In these selected HRTEM images with high image quality and their respective Bragg-filtered lattice points, no dislocations are detected near the $\alpha$/$\beta$-$\mathrm{PdH}_x$ phase boundary propagating along the [010] direction. While we cannot entirely exclude the possibility of rapidly moving dislocations at the interfaces, which may escape detection using our current experimental setup, the observations in \textbf{Fig. 3} suggest the presence of a coherent phase boundary within the hydriding 14 × 16 nm Pd nanocube.

The sequential \textit{d}-spacing maps (\textbf{Fig. 3c} ) uncover the detailed structural evolution of the hydrogen intercalation wave. As the wave propagates along the [010] direction during hydrogen intercalation, the initially non-planar and inclined $\alpha$/$\beta$-$\mathrm{PdH}_x$ interface undergoes a dynamic evolution, gradually aligning more with the (010) crystallographic plane. This structural rearrangement of the intercalation wave is visualized in a contour plot for the 14 × 16 nm Pd nanocube (\textbf{Fig. 3d} ). An analogous structural transition is observed in the 19 × 21 nm Pd nanocube during hydriding. The [100] propagating $\alpha$/$\beta$-$\mathrm{PdH}_x$ phase boundary progressively transforms and eventually closely aligns with the (100) plane (see \textbf{Extended Data} \textbf{Fig. 4}  for representative atomic-resolution HRTEM images and respective \textit{d}-spacing colormaps, and \textbf{Fig. 3e}  for a contour plot showing a temporal evolution of the hydrogen intercalation wave).
The dynamic reorientation of the hydrogen intercalation waves in Pd nanocubes manifests the system's continuous effort to reach a local mechanical equilibrium as the $\alpha$/$\beta$-$\mathrm{PdH}_x$ phase boundaries propagate. This provides critical insights into the atomic-level mechanisms driving hydrogen-induced phase transformations in Pd nanocrystals. Specifically, we propose that due to elastic interactions and the confined geometry of the nanocube, the most energetically favorable crystallographic orientation of the $\alpha$/$\beta$-$\mathrm{PdH}_x$ interface depends on the volume fraction (or composition) of each phase. For instance, based on the experimental results in \textbf{Fig. 3}, we hypothesize that when the volume fractions of $\alpha$- and $\beta$-$\mathrm{PdH}_x$ phases are approximately comparable, the \{100\} planes become the most stable interfacial orientation, which is corroborated by the Monte Carlo simulation results of the next section.

\subsection*{\label{sec:theory}Theoretical simulation of hydrogen intercalation wave mechanism}
To understand the basic thermodynamic driving forces underpinning the observed intercalation wave mechanism, we establish a minimal model to simulate the nanoscale dynamics of solid-solid phase transformations. Simulations of hydrogen absorption and desorption are conducted with kinetic Monte Carlo of the elastic Ising model Hamiltonian\cite{fratzl_modeling_1999, frechette_origin_2020, frechett_elastic_2021} (see \textbf{Supplementary Notes 3.1-3.3} and \textbf{Supplementary Fig. 8}). The elastic Ising model accounts for the energetic cost of the strain that is induced by the large difference in lattice constants between regions rich in $\alpha$-$\mathrm{PdH}_x$ and those rich in $\beta$-$\mathrm{PdH}_x$ within a nanocube.
In the model, the nanocube is coarse-grained so that each unit cell is defined as either the $\alpha$-$\mathrm{PdH}_x$ or $\beta$-$\mathrm{PdH}_x$ phase. The Hamiltonian is parameterized with experimental values of the elastic moduli of $\mathrm{PdH}_x$\cite{hsu_elastic_1979} (see \textbf{Supplementary Note 3.1} and \textbf{Supplementary Table 1}), enabling us to describe an effective interaction between sites mediated by elasticity. With this parameterization, the simple model reproduces a phase diagram (\textbf{Supplementary Fig. 9} ) that is quantitatively analogous to the experimental bulk phase diagram\cite{flanagan_palladium-hydrogen_1991} (see \textbf{Supplementary Note 3.4}). It also captures the expected effects of nanoscale confinement, in a suppressed critical temperature and an enhanced equilibrium concentration of hydrogen at the interface in the $\alpha$-$\mathrm{PdH}_x$ phase for smaller cubes\cite{yamauchi_nanosize_2008}.
To investigate the pathways of reversible absorption in our model, free energy calculations are carried out as a function of the $\beta$-$\mathrm{PdH}_x$ concentration ($c_\beta$) at coexistence (\textbf{Fig. 4a}) and away from it (\textbf{Fig. 4b}), using umbrella sampling (see \textbf{Supplementary Note 4}  and \textbf{Supplementary Table 2}). Representative configurations (see \textbf{Supplementary Note 5}) from four points along the reversible pathway (\textbf{Figs. 4c and 4d}  and \textbf{Supplementary Video 8}) recapitulate the experimentally identified hydrogen intercalation wave mechanism: nucleation at one corner followed by interface propagation along the $\langle 100\rangle$ directions. Specifically, \textbf{Fig. 4e}  shows that the corner-nucleated configuration features an interface, approximately along the $\langle 110\rangle$ direction, spanning the cube’s length. This $\alpha$/$\beta$-$\mathrm{PdH}_x$ interface then moves in either of the two equivalent $\langle 100\rangle$ directions, to join the opposite face of the cube, forming a stable \{100\}-type interface across the cube. In the reversible limit, transformations of hydrogen absorption and desorption in Pd nanocubes can be understood as proceeding through the same pathways, mediated by the same thermodynamic driving forces.

\textbf{Fig. 4b}  depicts the shift in the free energy curve from \textbf{Fig. 4a}  in response to a chemical potential difference, $\Delta$$\mu$, between the $\mathrm{H}_2$ molecules in solution and the H atoms in the palladium lattice, which would act as a chemical driving force for absorption. The curves show that the single-corner configuration represents a critical nucleus when the chemical driving force is significantly smaller than the thermal energy. Overall, the similarities between the simulated reversible pathway and the observed propagation dynamics, validate the hypothesis that across the range of sizes considered, the nanocubes are able to maintain a local mechanical equilibrium during the phase transformation and are not limited by kinetic factors, such as diffusion.
To further elucidate the absorption pathway, we study the elastic energies of idealized configurations along potential absorption pathways. We consider perfect interfaces moving across the cube in the $\langle 100\rangle$, $\langle 110\rangle$, and $\langle 111\rangle$ directions. In \textbf{Extended Data Fig. 5} , we compute the energies for different nanocube sizes for each of these interfaces. Our results show that while the energetic cost to form any interface scales extensively with system size, for each size, the $\langle 100\rangle$ interface is significantly energetically preferred near $c_\beta = 0.5$. However, for $c_\beta$ below $0.2$, the most energetically favorable interface is along the $\langle 110\rangle$ direction, corresponding to the corner-nucleated configurations demonstrated in both the model and experiments.

Our model does not include effects from $\mathrm{H}_2$ diffusion in solution or defects in the nanocube. This indicates that the observed absorption pathway and structural evolution of the hydrogen intercalation wave can be understood predominantly in terms of minimizing the elastic energy cost associated with the interface between the two phases at each $c_\beta$ value. This finding deepens our fundamental understanding of hydrogen intercalation processes in palladium nanocrystals. Moreover, the relative energies among the $\langle 100\rangle$, $\langle 110\rangle$, and $\langle 111\rangle$ directions along the sorption pathway do not change across a broad range of system sizes. Explicit free energy curves are presented only for an approximately 6 nm cube due to computational constraints (\textbf{Fig. 4}), but because the energetic ordering of the facets remains consistent for nanoparticle sizes up to 14 nm (\textbf{Extended Data Fig. 5a}), we should expect that the simulated reversible pathway applies consistently to all the nanocubes studied in our liquid-phase TEM experiments.

Notably, our simple model successfully reproduces the experimental observations based solely on the elastic moduli of the material. Given its minimal ingredients, this model is applicable to a broad class of systems where solid-solid phase transformations involve changes in local composition and the generation of large strains. Indeed, this minimal model has been employed to study two-component alloys with elastic misfit and cation-exchange processes in quantum dots\cite{frechett_elastic_2021, cahn_simple_1984, fratzl_ising_1996, gupta_microscopic_2001}. We anticipate that, with the appropriate parameterization, it could be similarly applied to lithium intercalation systems like iron phosphate nanoparticles and semiconducting bilayers. Thus, the insights we gain concerning phase transformations in nanoparticles with this combination of interaction type, strength, and range allow us to uncover universal behaviors that should, in principle, be observable across a broad spectrum of material systems.

\section*{\label{sec:conc}Conclusion}
Utilizing liquid-phase TEM, we have achieved unprecedented atomic-resolution imaging of hydrogen (de)intercalation dynamics in palladium nanocrystals down to the unexplored size regime ($\smallsim$10 nm). Our observations reveal that the intercalation wave mechanism governs hydrogen-induced phase transformations in Pd nanocubes ranging from $\smallsim$60 to $\smallsim$10 nm, rather than transitioning to the shrinking-core mechanism at smaller sizes. We further discover the evolution of hydrogen intercalation wave dynamics at the atomic level: as the phase transformations proceed, non-planar and inclined $\alpha$/$\beta$-$\mathrm{PdH}_x$ interfaces change to become more aligned with \{100\} planes. Employing an elastic Ising model, we primarily explain these intercalation wave dynamics through the continuous minimization of elastic energies at the $\alpha$/$\beta$-$\mathrm{PdH}_x$ interfaces.

Our discoveries provide important insights for advancing intercalation-based devices such as batteries, hydrogen storage, and neuromorphic computing devices. Direct visualization of phase morphologies not only guides the interpretation of experimental results but also refines theoretical models predicting material behaviors during solute-induced phase transformations. The insights gained from this study may open opportunities to design intercalation materials that undergo desired phase transformations with tailored intercalation pathways. This approach can help suppress mechanical deformation during solute (de)intercalation, thereby improving the performance and lifespan of hydrogen storage systems, batteries, and other intercalation-mediated devices. Our ability to directly visualize atomic-scale dynamics opens future opportunities for identifying undiscovered intercalation dynamics and metastable phases across a wide spectrum of material systems.

\begin{figure*}[!htbp]
    \includegraphics[width=0.75\textwidth]{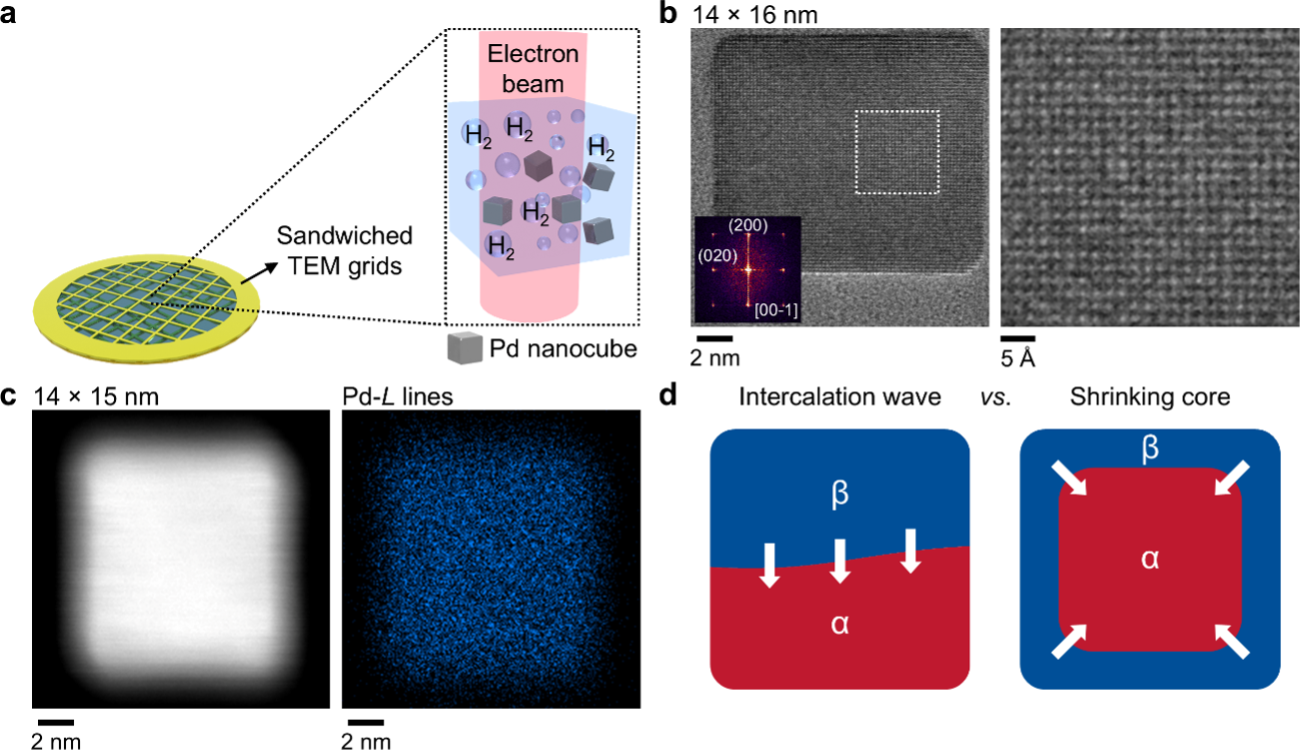}
    \caption{\textbf{In situ liquid-phase TEM investigation of hydrogen-induced phase transformations in Pd nanocubes.} \textbf{a}, Schematic illustration of the experimental setup, where the electron beam facilitates in situ hydrogen generation for absorption studies. Under our experimental conditions, radiolysis of the aqueous solution predominantly produces hydrogen and oxygen gases. \textbf{b}, HRTEM image (left) of a palladium nanocube (14 × 16 nm), captured as a frame from in situ liquid-phase TEM data during hydrogen absorption (\textbf{Supplementary Video 6}). The respective FFT pattern exhibits (200) and (020) reflections along the [00-1] zone axis, with the nanocube in the $\alpha$-$\mathrm{PdH}_x$ phase. An enlarged HRTEM image (right), highlighting the selected region (dotted area), demonstrates atomic resolution achieved directly from the in situ movie data. \textbf{c}, STEM image (left) and corresponding Pd-L lines EDS map (right) of a representative Pd nanocube (14 × 15 nm) utilized in this liquid-phase TEM study. \textbf{d}, Schematic illustration showing two representative morphologies of phase separation in solute intercalation materials. Here, hydrogen absorption in Pd nanocubes with $\alpha$- and $\beta$-$\mathrm{PdH}_x$ phases is used as an example to depict these two mechanisms.}
    \label{fig:1}
\end{figure*}

\begin{figure*}[!htbp]
    \includegraphics[width=0.75\textwidth]{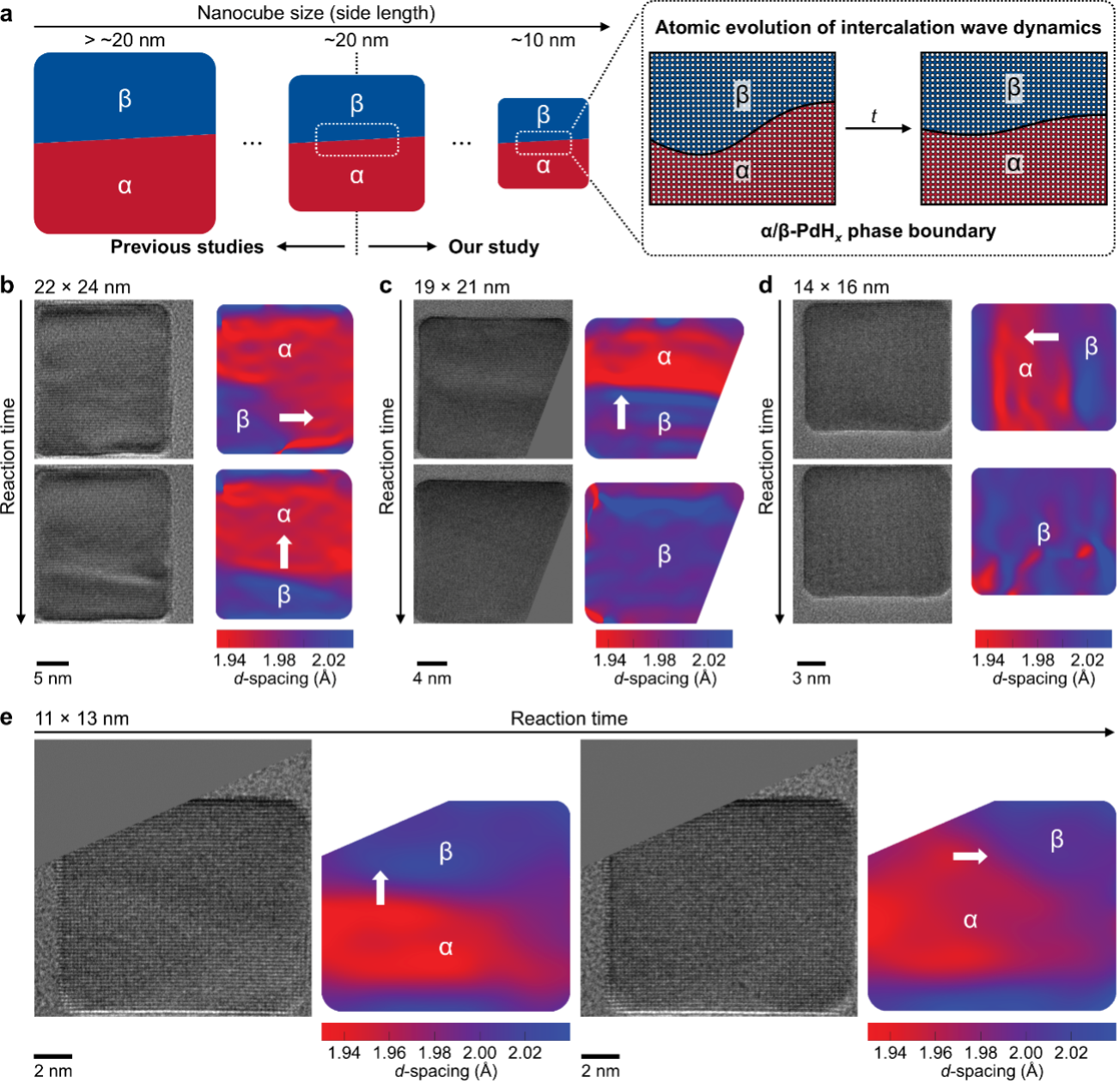}
    \caption{\textbf{In situ atomic-resolution observation of phase transformation pathways and }$\mathbf{\alpha}$/$\mathbf{\beta}$-$\mathbf{\mathrm{PdH}_x}$ \textbf{interface structures in Pd nanocubes during hydrogen-induced phase transformations.} \textbf{a}, Illustration (left) comparing the nanocube size ranges explored in previous studies with those in this study, extending the size limit down to $\smallsim$10 nm. Schematic (right) depicting the newly observed phenomenon: the dynamic evolution of hydrogen intercalation waves ($\alpha$/$\beta$-$\mathrm{PdH}_x$ interfaces) in Pd nanocubes, transitioning from non-planar and inclined boundaries to those more closely aligned with \{100\} planes, as they propagate along the $\langle 100\rangle$ directions. \textbf{b-e}, Sequential HRTEM images and corresponding (200) or (020) \textit{d}-spacing colormaps of Pd nanocubes (11 to 24 nm) during hydrogen absorption (\textbf{b-d}) and desorption (\textbf{e}). These HRTEM images are extracted from \textbf{Supplementary Videos 4-7}. Arrows indicate the propagation directions of the $\alpha$/$\beta$-$\mathrm{PdH}_x$ interfaces. The results corroborate that the intercalation wave mechanism, instead of the shrinking-core mechanism, is consistently observed across Pd nanocubes of varying sizes from $\smallsim$10 to $\smallsim$60 nm during both absorption and desorption.
}
    \label{fig:2}
\end{figure*}
\begin{figure*}[!htbp]
    \includegraphics[width=0.50\textwidth]{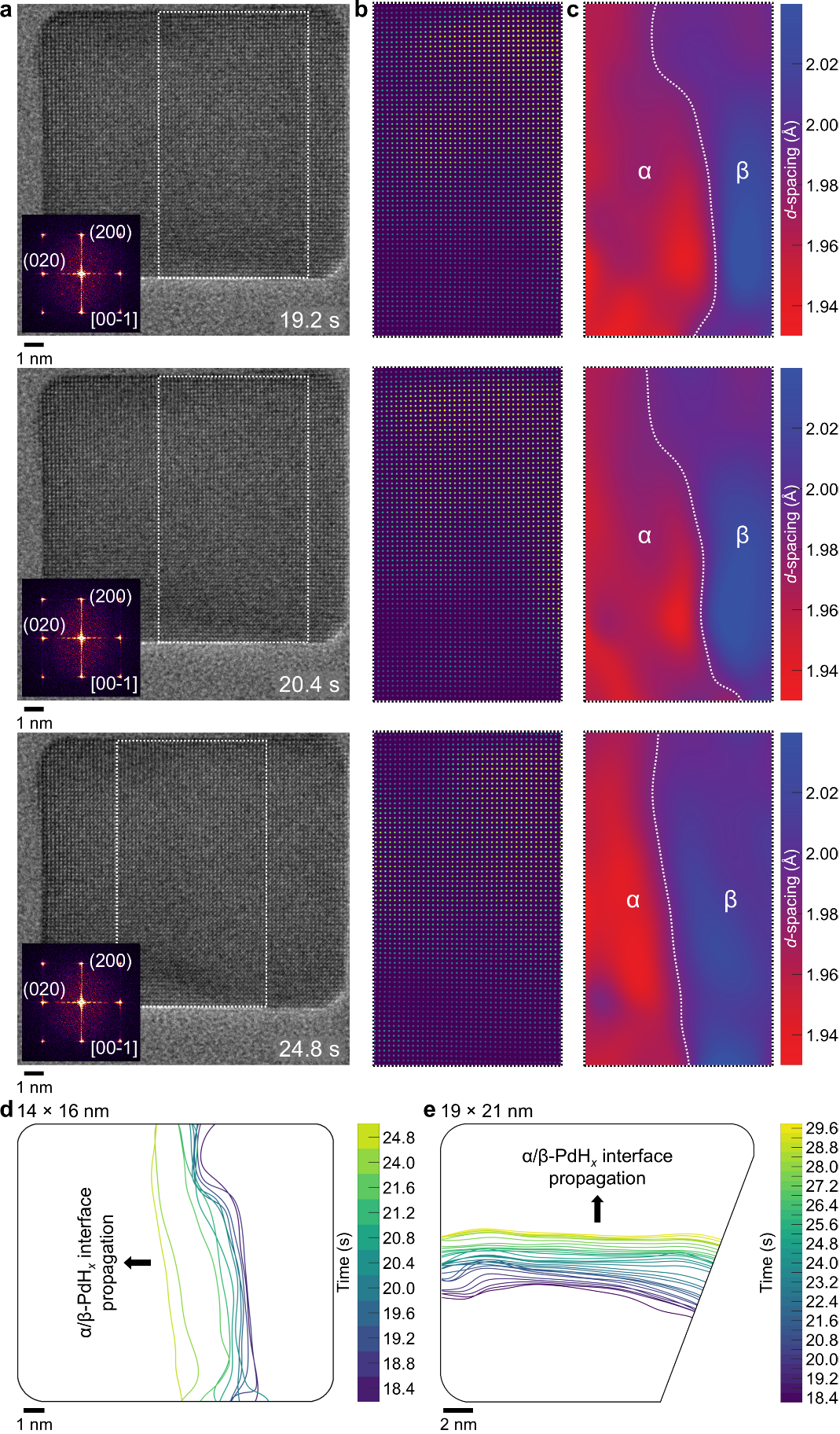}
    \caption{\textbf{Atomic evolution of hydrogen intercalation waves (i.e.,} $\mathbf{\alpha}$/$\mathbf{\beta}$-$\mathbf{\mathrm{PdH}_x}$ \textbf{phase boundaries) propagating along $\langle 100\rangle$ directions during hydrogen absorption in Pd nanocrystals sized 14 × 16 nm and 19 × 21 nm (based on Supplementary Videos 6 and 5, respectively).} \textbf{a},\textbf{b},\textbf{c}, (Top to bottom) Sequential HRTEM images with their respective FFT patterns (\textbf{a}), demonstrating atomic-resolution visualization of hydrogen absorption in the 14 × 16 nm Pd nanocrystal. Bragg-filtered lattice points derived from (200) and (020) lattice planes (see \textbf{Supplementary Fig. 7} ) (\textbf{b}) and (020) \textit{d}-spacing colormaps (\textbf{c}), corresponding to the white-dotted regions in (\textbf{a}). Each row represents a single frame. \textbf{d},\textbf{e}, Contour plots illustrating the temporal evolution of $\alpha$/$\beta$-$\mathrm{PdH}_x$ interfaces during hydrogen absorption in Pd nanocubes sized 14 × 16 nm (\textbf{d}) and 19 × 21 nm (\textbf{e}). Black arrows indicate the propagation directions of the $\alpha$/$\beta$-$\mathrm{PdH}_x$ interfaces. In (\textbf{c-e}), the white-dotted lines and contours trace the $\alpha$/$\beta$-$\mathrm{PdH}_x$ interfaces, determined by the average (200) \textit{d}-spacing value, between the smallest (regarded as $\alpha$-$\mathrm{PdH}_x$) and largest ($\beta$-$\mathrm{PdH}_x$) experimentally measured values for similarly sized nanocubes\cite{shi_solution-phase_2022}.
}
    \label{fig:3}
\end{figure*}
\begin{figure*}[!htbp]
    \includegraphics[width=0.75\textwidth]{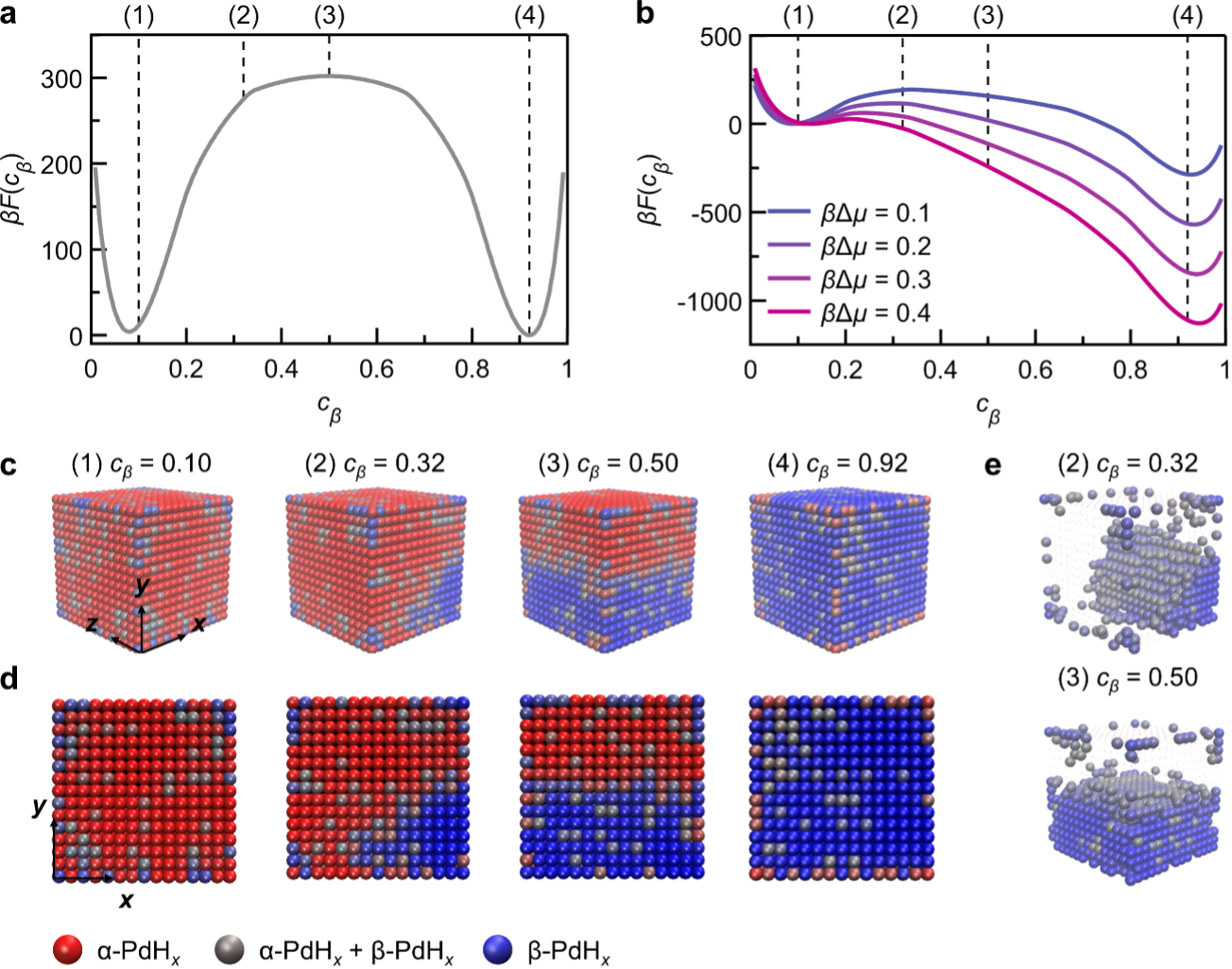}
    \caption{\textbf{Free energy curves for hydrogen absorption and configurations along a reversible absorption pathway in the model.} \textbf{a}, The Helmholtz free energy as a function of $c_\beta$ computed from standard umbrella sampling techniques and the weighted histogram analysis method. All simulations are run at temperature (T) = 300 K (complete simulation details in \textbf{Supplementary Note 4} ). \textbf{b}, The same free energy curves as in (\textbf{a}), but with linear shifts due to chemical potential differences, $\Delta$$\mu$, between $\mathrm{H}_2$ molecules in the surrounding solution and H atoms in the palladium lattice. One sees the corner nucleated configuration, (2), is a representative critical nucleus when chemical driving for absorption is on the order of one-tenth the thermal energy. Note that, when multiplying an energy, $\beta$ represents the inverse temperature. \textbf{c},\textbf{d}, Two perspectives on representative configurations taken from umbrella sampling trajectories, with $c_\beta$ increasing in each configuration from left to right. \textbf{e}, The same configurations, (2) and (3) from (\textbf{c} and \textbf{d}), where only unit cells with a local value of $c_\beta$ greater than 0.5 are shown to visualize the full interphase boundaries.
}
    \label{fig:4}
\end{figure*}

\cleardoublepage

\section*{\label{sec:methods}Methods}
\subsection*{\label{syn}Synthesis of Pd nanocubes}
We synthesize Pd nanocubes with three representative size ranges: $\smallsim$10 to $\smallsim$20 nm, $\smallsim$20 to $\smallsim$30 nm, and $\smallsim$40 to $\smallsim$60 nm, based on the synthesis protocols in the cited publications\cite{sytwu_visualizing_2018, jin_synthesis_2011}. For the $\smallsim$10 to $\smallsim$20 nm nanocubes, an aqueous solution is prepared by dissolving 52.5 mg of poly(vinyl pyrrolidone) (PVP, with an average molecular weight of $\smallsim$55,000, Sigma-Aldrich), 30 mg of L-ascorbic acid (Sigma-Aldrich), and 150 mg of KBr (Sigma-Aldrich) in 4 mL of deionized (DI) water. This solution is preheated at 80 °C for 10 minutes under magnetic stirring. Following this, 1.5 mL of another aqueous solution containing 28.5 mg of $\mathrm{Na}_2\mathrm{PdCl}_4$ (Acros Organics) is injected in one shot into the preheated solution. The reaction mixture is kept under stirring at 80 °C for 3 hours. Afterward, the nanocubes are collected by centrifugation and washed three times with DI water. For the synthesis of $\smallsim$20 to $\smallsim$30 nm Pd nanocubes, the steps are identical to those for the $\smallsim$10 to $\smallsim$20 nm nanocubes, except the 150 mg of KBr is replaced with 1000 mg of KBr.
For the synthesis of Pd nanocubes in the size range of $\smallsim$40 to $\smallsim$60 nm, a two-step procedure is employed. In the first step, the seed solution is prepared by dissolving 91.1 mg of cetrimonium bromide (CTAB, Sigma-Aldrich) in 20 mL of DI water and stirring it at 95 °C for 5 minutes. A 10 mM $\mathrm{H}_2\mathrm{PdCl}_4$ solution is generated by dissolving 8.9 mg of $\mathrm{PdCl}_2$ (Sigma-Aldrich) in 5 mL of 20 mM HCl (Sigma-Aldrich). Then, 1 mL of this 10 mM $\mathrm{H}_2\mathrm{PdCl}_4$ solution is added to the CTAB solution, which turns red. The mixture is stirred for 5 minutes before 160 $\mu$L of a 100 mM L-ascorbic acid solution (produced by dissolving 17.6 mg of L-ascorbic acid in 1 mL of DI water) is added. The final mixture is stirred for 5 minutes at 95 °C, and then placed in an ice bath to stop the reaction. In the second step, the size of the Pd nanocubes is increased from the seed solution to the desired $\smallsim$40 to $\smallsim$60 nm range. A 0.1 M CTAB solution is formed by dissolving 364.5 mg of CTAB in 10 mL of DI water and heating it to 60 °C. Then, 125 $\mu$L of the 10 mM $\mathrm{H}_2$PdCl4 solution, 100 $\mu$L of the seed solution, and 50 $\mu$L of the 100 mM L-ascorbic acid solution are added to this CTAB solution. The mixture is slightly shaken and allowed to sit for 2 hours at 60 °C. The final product is collected by centrifugation and washed twice with DI water.
\subsection*{\label{fab}Liquid cell fabrication}
In our liquid-phase TEM experiments, liquid cells are constructed using TEM grids coated with two types of electron-transparent membranes: ultrathin carbon films (3-4 nm thick, Electron Microscopy Sciences) and Formvar films (14-20 nm thick). For Pd nanocubes $\smallsim$20 nm and smaller, we use liquid cells with ultrathin carbon membranes, and Pd nanocubes larger than $\smallsim$20 nm are studied using Formvar-based liquid cells. The Formvar-coated TEM grids are fabricated through the following process: a glass microscope slide is first dipped into a 0.25\% Formvar solution in ethylene dichloride (Electron Microscopy Sciences), forming a thin film of Formvar on the slide (\textbf{Supplementary} \textbf{Fig. 2a}). This film is then air-dried for 5 minutes. To separate the Formvar film from the glass slide, the slide is immersed in water, and the surface tension lifts the Formvar film, which floats on the water's surface (\textbf{Supplementary} \textbf{Fig. 2b}). TEM grids (Electron Microscopy Sciences) are gently placed onto the floating film (\textbf{Supplementary} \textbf{Fig. 2c}). Finally, the Formvar-coated TEM grids are retrieved using another glass slide and left to dry under ambient conditions (\textbf{Supplementary} \textbf{Fig. 2d}). Using electron energy loss spectroscopy, the thickness of the fabricated Formvar films on TEM grids is consistently measured to be between 14 and 20 nm.
To fabricate a liquid cell, we first plasma-clean a TEM grid using a Model 1020 Plasma Cleaner (Fischione Instruments) with a mixture of argon and oxygen gases. This step makes the electron-transparent membranes hydrophilic and gently removes organic contamination from TEM grids. Next, 1-2 $\mu$L of an aqueous dispersion containing Pd nanocubes is drop-cast onto the plasma-treated TEM grid and left to air-dry. A second pristine TEM grid is prepared, and both the nanocube-loaded and pristine TEM grids are plasma-cleaned. Then, a droplet ($\leq$ 0.5 $\mu$L) of an aqueous KOH (Sigma-Aldrich) solution is cast onto one of the plasma-cleaned TEM grids. If needed, filter paper is used to carefully remove excess KOH solution. Empirically, under our experimental conditions, we observe that a small amount of lower concentration KOH solution decreases the electron-beam-induced generation of radiolytic products, including hydrogen gas. Consequently, atomic-resolution HRTEM imaging is carried out using 0.01 M KOH aqueous solution, and other experimental results are based on 0.1 or 0.05 M KOH solution. The second plasma-treated TEM grid is gently placed on top of the droplet-loaded TEM grid. Van der Waals forces between the membranes seal the liquid cell and encapsulate a thin liquid film between the two sandwiched grids. The assembled liquid cell is subsequently loaded into a transmission electron microscope for liquid-phase TEM experiments.
\subsection*{\label{tem}Transmission electron microscopy}
The in situ liquid-phase TEM and ex situ S/TEM experiments are conducted using a FEI ThemIS 60-300 transmission electron microscope (Thermo Fisher Scientific), equipped with an extreme field emission gun (X-FEG) operated at accelerating voltages of 200 and 300 kV. This microscope features an image aberration corrector, allowing for a spatial resolution limit of 70 pm at 300 kV. It is also equipped with a SuperX EDS detector, incorporating four windowless silicon drift detectors, and a Ceta2 complementary metal-oxide semiconductor (CMOS) camera with a scintillator. Atomic-resolution HRTEM images are acquired with the aid of the AXON Synchronicity in situ TEM software (Protochips). Additionally, electron energy loss spectroscopy data are collected using an F20 UT Tecnai transmission electron microscope (Thermo Fisher Scientific).
\subsection*{\label{img}Image processing and analysis}
For the lower magnification TEM videos (\textbf{Supplementary Videos 2-3}), two consecutive TEM images are averaged to create each frame. To generate each frame of the atomic-resolution HRTEM videos (\textbf{Supplementary Videos 4-7}), we average two successive HRTEM images and subsequently carry out the non-linear Fourier filtering of each averaged frame to reduce Gaussian noise (\textbf{Supplementary Fig. 6}). The two-dimensional (2D) Fourier transform of each averaged HRTEM image is computed while preserving the phase angle information. The magnitude of the Fourier transform is adjusted by applying an absolute value shrinkage filter, which varies linearly with spatial frequency in the Fourier domain. This non-linear filtering process subtracts the filter from the magnitude and sets any resultant negative values to zero. Eventually, the inverse Fourier transform is applied using the modified magnitude and preserved phase angle information to reconstruct the filtered HRTEM image.
The \textit{d}-spacing colormaps and Bragg-filtered images corresponding to the filtered HRTEM images are generated using geometric phase analysis\cite{hytch_quantitative_1998} (GPA). First, in the Fourier domain, we select a pair of Bragg peaks of interest and measure the position of one of these Bragg peaks, which serves as the reference reciprocal lattice vector, $\mathbf{g}$. We generate a 2D Bragg filtering mask centered on this Bragg peak, using a flat-topped circular function with a smooth decay to zero near its edge. The 2D Fourier transform of the HRTEM image is then computed, and the Bragg filter is applied. The inverse Fourier transform of the filtered Fourier transform results in the complex image, $H'_g(\mathbf{r})$, where r represents the spatial coordinates. The Bragg-filtered image, $B_g(\mathbf{r})$, is derived from the real part of $H'_g(\mathbf{r})$. We also obtain the phase image, $P_g(\mathbf{r})$, by subtracting $2\pi\mathbf{g}\cdot\mathbf{r}$ from the raw-phase image of the $H'_g(\mathbf{r})$. Taking the spatial derivative of this phase image, we compute the strain map, $\epsilon_g(\mathbf{r})$, with respect to the $\mathbf{g}$ vector. Finally, we convert the strain map into the \textit{d}-spacing map using the \textit{d}-spacing value corresponding to the reference lattice vector, $\mathbf{g}$.

\begin{acknowledgments}
This work was supported by the U.S. Department of Energy, Office of Science, Office of Basic Energy Sciences (BES), Materials Sciences and Engineering Division under Contract No. DE-AC02-05-CH11231 within the in-situ TEM program (KC22ZH). Work at the Molecular Foundry of Lawrence Berkeley National Laboratory (LBNL) was supported by the U.S. Department of Energy under Contract No. DE-AC02-05CH11231. D.L. acknowledges the Kwanjeong Study Abroad Scholarship from the KEF (Kwanjeong Educational Foundation) (KEF-2019). S.B.B. thanks the Alexander von Humboldt Foundation for financial support and Prof. A. Paul Alivisatos for discussions. We appreciate Dr. Karen C. Bustillo and Dr. Rohan Dhall for their assistance with the electron microscope setup.
\end{acknowledgments}

\titlespacing*{\subsection} {0pt}{3ex}{0ex}

\subsection*{Author Contributions}
D.L. and S.B.B. conducted the in situ liquid-phase TEM and ex situ S/TEM experiments. S.B.B., Y.S., and S.Z. synthesized the Pd nanocubes. D.L., S.B.B., and C.O. carried out the experimental data analysis. S.O.-L. performed the theoretical modeling and simulations and analyzed the results under the supervision of D.T.L. This paper was written by D.L., S.O.-L., S.B.B., D.T.L., and H.Z., with contributions from all co-authors. H.Z. conceived and supervised the project.
\subsection*{Competing Interests}
Authors declare that they have no competing interests.
\subsection*{Correspondence} 
Correspondance and requests for materials should be addressed to H. Zheng  (hmzheng@lbl.gov) or D. T. Limmer (dlimmer@berkeley.edu).
\titleformat*{\section}{\Large\bfseries}

\subsection*{\label{sec:data}Data and Code Availability}
All data and codes that support the findings of this study are available from the corresponding authors upon reasonable request.

\cleardoublepage

\begin{widetext}
\section*{Extended Data}
\begin{figure}[!htbp]
    \includegraphics[width=0.75\textwidth]{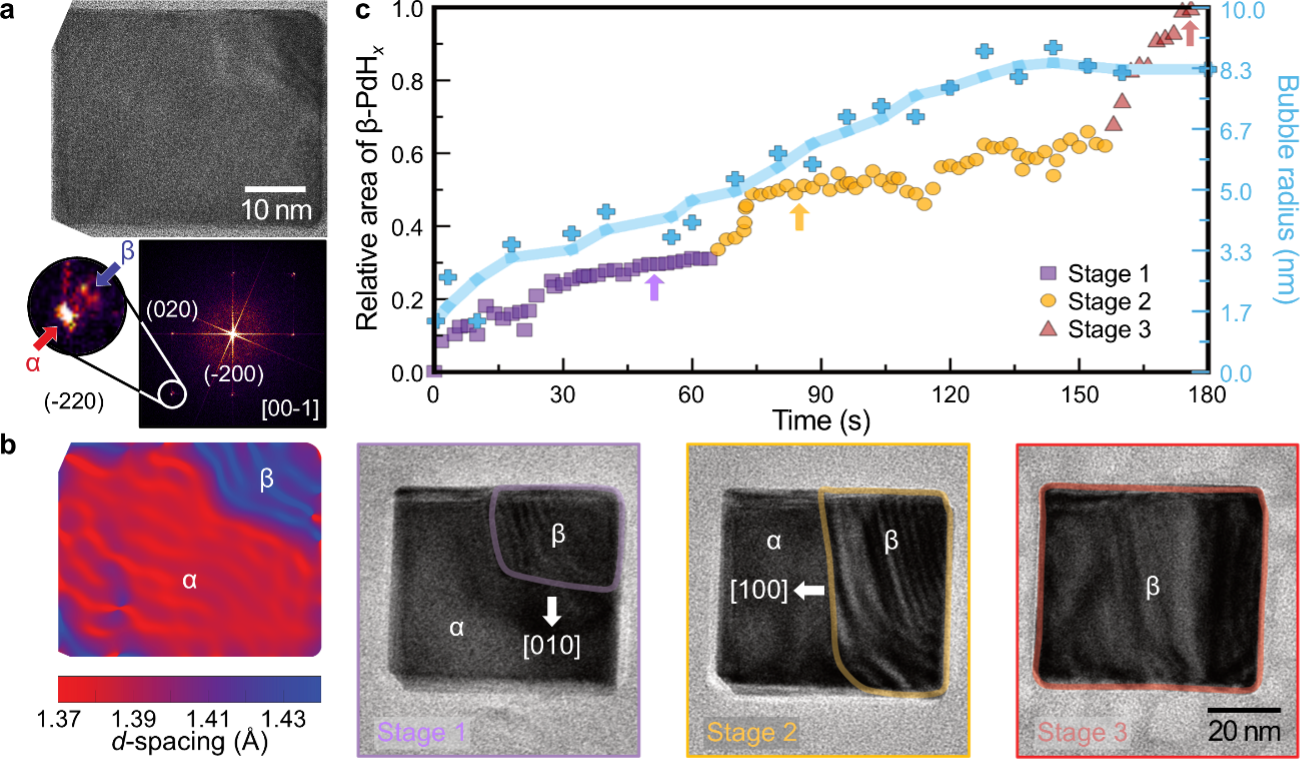}
    \caption*{\textbf{Extended Data Fig.\ 1: Hydrogen intercalation wave mechanisms during hydrogen absorption in Pd nanocubes in the $\smallsim$20 to $\smallsim$60 nm size range.} \textbf{a}, TEM image of a Pd nanocube (36 × 45 nm) at the initial state of the hydrogen absorption reaction. The respective FFT pattern shows the presence of two crystal phases ($\alpha$ and $\beta$ phases) as indicated by a splitting of the (-220) reflection spots. \textbf{b}, The analysis of the spatial variation of the (-220) \textit{d}-spacing reveals the local lattice expansion at the upper right corner of the nanocube. The local lattice expansion induces strain and a local thickness variation, resulting in the dark fringes in the respective TEM image. \textbf{c}, Phase transformation dynamics (\textbf{Supplementary Video 1}) determined based on the relative area of a nanocube (55 × 61 nm) covered by the dark fringes (i.e., the relative area of the $\beta$-$\mathrm{PdH}_x$) and the bubble radius, which is proportional to hydrogen concentration. The reaction happens in three stages, highlighted in purple, yellow, and red. The displayed TEM images are characteristic of the nanocube during the three stages (designated by colored arrows in the plot). The experiment time is given relative to the initial data acquisition.}
    \label{fig:ex1}
\end{figure}
\begin{figure}[!htbp]
    \includegraphics[width=0.75\textwidth]{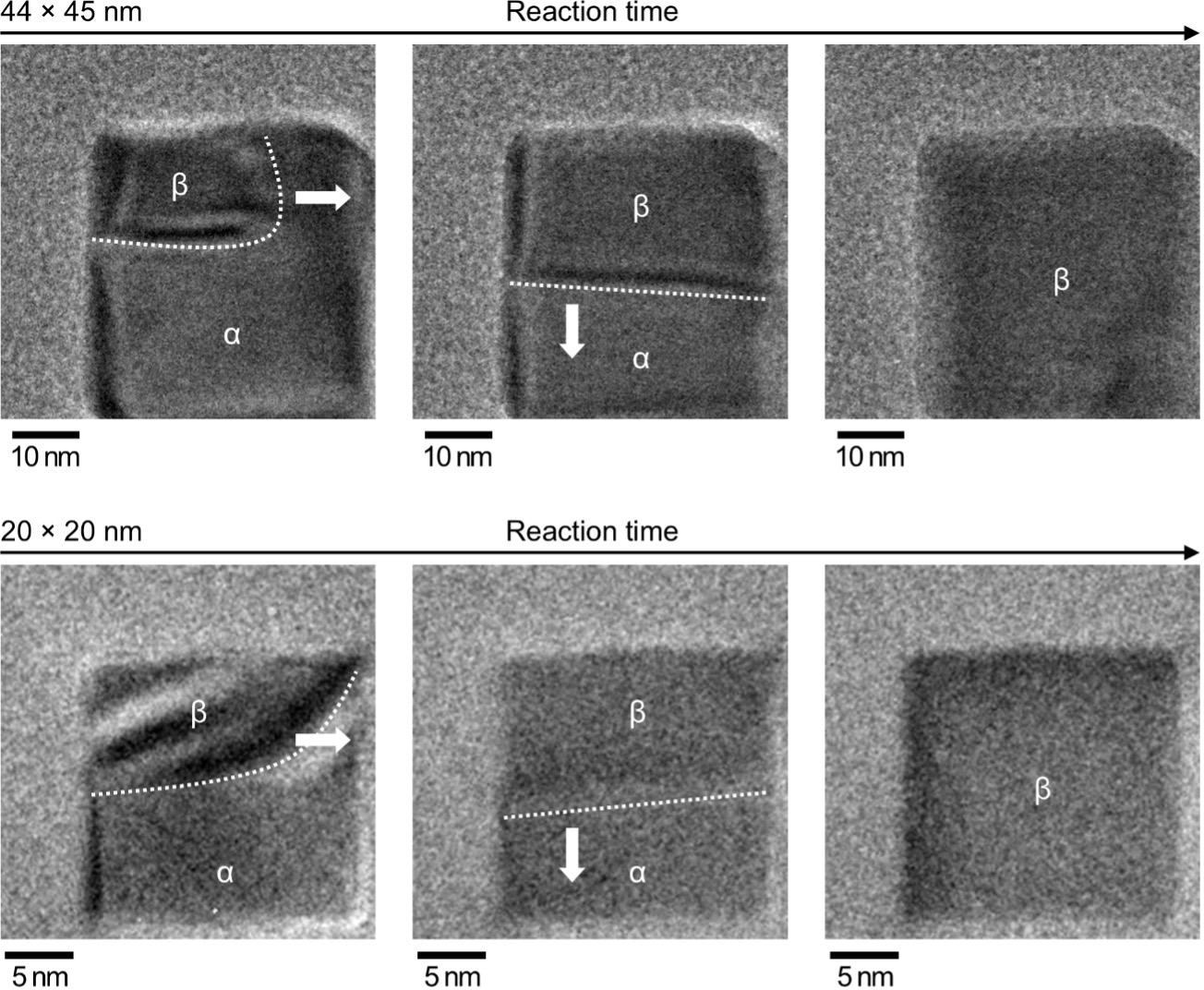}
    \caption*{\textbf{Extended Data Fig.\ 2: Additional evidence of hydrogen intercalation wave mechanisms during hydrogen absorption in Pd nanocubes in the $\smallsim$20 to $\smallsim$60 nm size range.} Time-ordered TEM images of Pd nanocubes with sizes ranging from 20 to 45 nm. Each row represents Pd nanocubes of different sizes and illustrates key steps of the hydrogen absorption process: corner nucleation, expansion to one of the \{100\} facets, and subsequent $\langle 100\rangle$ propagation. Dark fringes, attributed to the nucleated $\beta$-$\mathrm{PdH}_x$ phase, are used to follow the phase transformation pathways. Dotted lines denote the approximate locations of the phase boundaries, and arrows indicate the propagation direction of the $\alpha$/$\beta$-$\mathrm{PdH}_x$ interfaces. These TEM images are taken from \textbf{Supplementary Videos 2-3}.}
    \label{fig:ex2}
\end{figure}
\begin{figure}[!htbp]
    \includegraphics[width=0.60\textwidth]{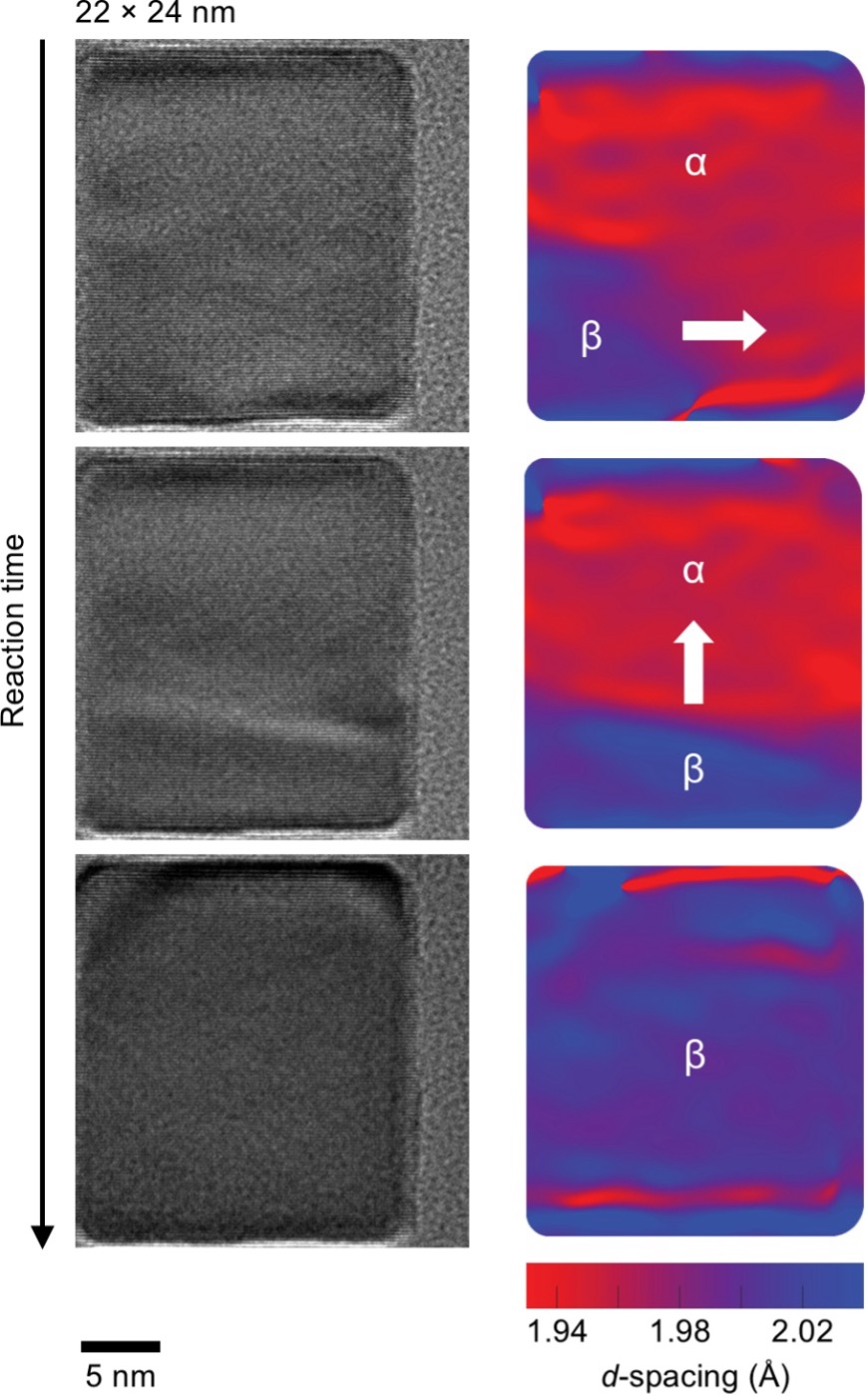}
    \caption*{\textbf{Extended Data Fig.\ 3: Complete set of sequential HRTEM images and corresponding (200) \textit{d}-spacing colormaps of the hydriding 22 × 24 nm Pd nanocube (same particle as shown in \textbf{Fig. 2b} and \textbf{Supplementary Video 4)}.} These images and colormaps demonstrate the full hydriding process from $\alpha$- to $\beta$-$\mathrm{PdH}_x$ through the hydrogen intercalation wave propagation.}
    \label{fig:ex3}
\end{figure}
\begin{figure}[!htbp]
    \includegraphics[width=0.75\textwidth]{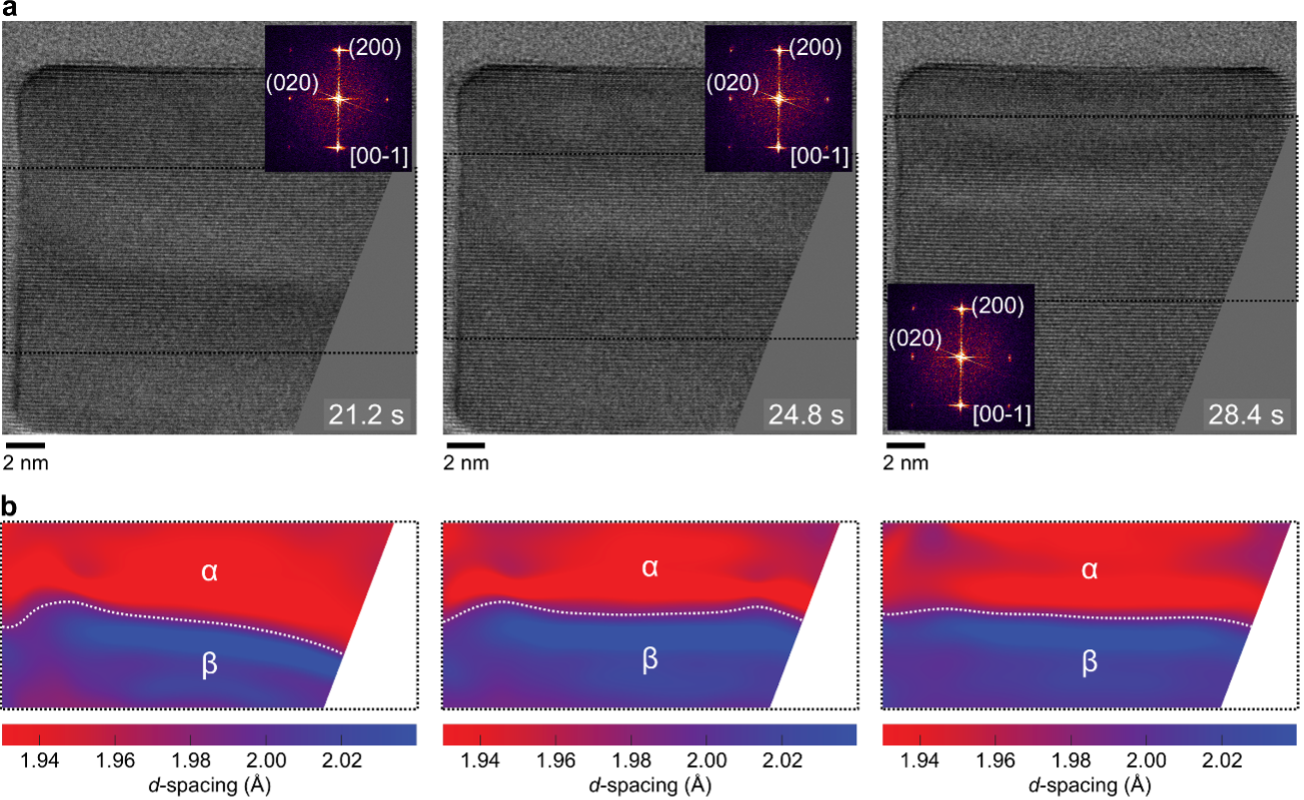}
    \caption*{\textbf{Extended Data Fig.\ 4: Evolution of the $\mathbf{\alpha}$/$\mathbf{\beta}$-$\mathbf{\mathrm{PdH}_x}$ phase boundary propagating along the [100] direction during hydrogen absorption in the 19 × 21 nm Pd nanocrystal (same particle as shown in \textbf{Fig. 2c}, \textbf{Fig. 3e} , and \textbf{Supplementary Video 5)}.} \textbf{a}, Sequential HRTEM images with corresponding FFT patterns of the 19 × 21 nm Pd nanocube. \textbf{b}, (200) \textit{d}-spacing colormaps of the black-dotted regions from the HRTEM images in (a), depicting the interface between $\alpha$- and $\beta$-$\mathrm{PdH}_x$ regions. Each column represents a single frame. In (\textbf{b}), white-dotted lines mark the $\alpha$/$\beta$-$\mathrm{PdH}_x$ interfaces, determined using the same criteria as in \textbf{Fig. 3.}}
    \label{fig:ex4}
\end{figure}
\begin{figure}[!htbp]
    \includegraphics[width=0.75\textwidth]{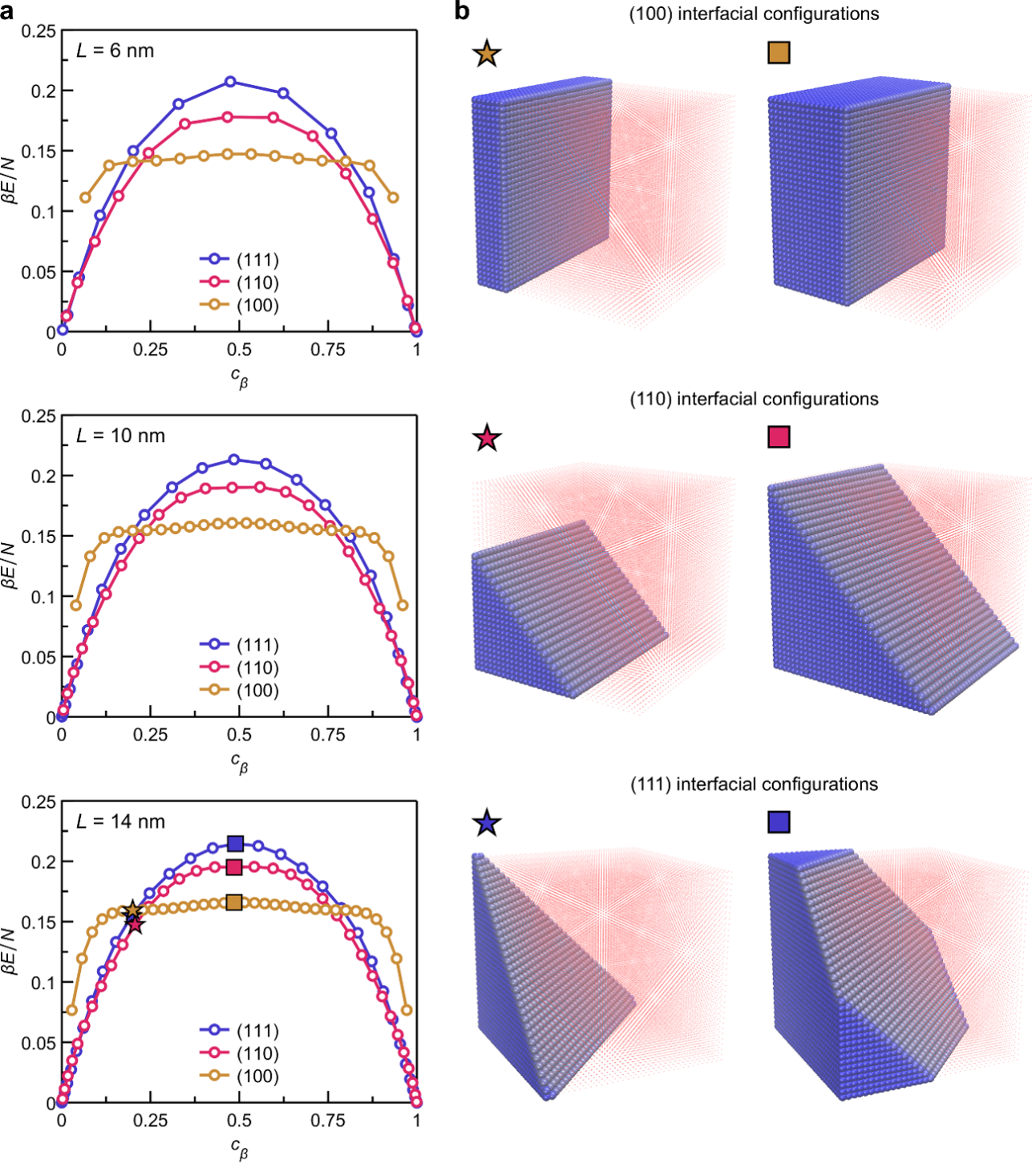}
    \caption*{\textbf{Extended Data Fig.\ 5:  An energetic analysis of idealized configurations along hypothetical absorption pathways.} \textbf{a}, The elastic energies (\textbf{Supplementary Equation (2)} in \textbf{Supplementary Note 3.2} ) of idealized (100), (110), and (111) interfacial configurations as a function of $c_\beta$, at three different sizes of nanocube. $N$ is the total number of sites in each the cube, $L$ is the side length of the nanocube, and $\beta$ represents the inverse temperature when multiplying an energy. \textbf{b}, Selected configurations with labels corresponding to the markers in the third panel of (\textbf{a}).}
    \label{fig:ex5}
\end{figure}

\cleardoublepage

\section*{Supplementary Notes}

\titleformat*{\section}{\normalsize\bfseries}
\section{Indirect estimation of $\mathrm{H}_2$ concentration as a function of electron dose rate}
During liquid-phase TEM experiments, nanobubbles form and grow when the concentrations of gases dissolved in the liquid exceed their saturation concentrations\cite{grogan_bubble_2014, kim_initial_2021}, a condition known as gas supersaturation. We track the bubble radius over time as a function of the electron dose rate to understand the $\mathrm{H}_2$ and $\mathrm{O}_2$ concentrations during in situ TEM experiments. An example TEM image (\textbf{Supplementary} \textbf{Fig. 3a}) highlights the identified gas bubbles. To account for the non-spherical bubbles, ellipses are fitted to the detected bubbles in the image. We estimate the bubble radius ($r$) using $r = (\mathbf{r}_1 \cdot \mathbf{r}_2)0.5$, where $\mathbf{r}_1$ and $\mathbf{r}_2$ are the minor and major radii of the detected ellipse, respectively (see \textbf{Supplementary} \textbf{Fig. 3b} ). The variation in bubble size is reflected in the broad distribution of bubble radii observed in a single TEM image. \textbf{Supplementary} \textbf{Fig. 3c}  shows the evolution of bubble radii at different time points during the in situ liquid-phase experiment. Considering that the sizes of the largest bubbles reflect the supersaturation of the solution, we track the average radius of the largest 25\% of bubbles as a function of the dose rate or illumination time (highlighted in red in \textbf{Supplementary} \textbf{Fig. 3c} ).
Our analysis demonstrates that the bubble radius reaches a steady-state size over reaction time at a given dose rate (\textbf{Supplementary} \textbf{Fig. 3d} ). The steady-state radius depends on the electron dose rate used during the experiments (\textbf{Supplementary Fig. 4} ), suggesting that the dose rate can be used to control the hydrogen concentration within the liquid cell, in line with previous studies\cite{grogan_bubble_2014, kim_initial_2021, schneider_electronwater_2014}. The time required to establish the steady-state concentrations varies randomly and does not follow the expected trend; for example, the higher the dose rate, the shorter the time. This is likely due to the different sizes of the investigated liquid films in the liquid cell. Since only a small subsection of the liquid cell is illuminated by the beam during liquid-phase TEM experiments, mass transport outside the illuminated region also influences the steady-state concentrations\cite{grogan_bubble_2014}.
\section{Hydrogen intercalation wave propagation dynamics at lower magnification}
\textbf{Extended Data} \textbf{Fig. 1c} presents the phase transformation dynamics during hydrogen absorption in the 55 × 61 nm Pd nanocube, based on the proportion of the area covered by dark fringes. As the main text describes, these dark fringes correspond to the $\beta$-$\mathrm{PdH}_x$ phase induced by hydrogen absorption. During this in situ liquid-phase experiment, the electron flux was stepwise increased to raise the hydrogen concentration, which can correlate with the average bubble radius (\textbf{Supplementary Note 1}). Combining time-series data of both the $\beta$-$\mathrm{PdH}_x$ proportion and the average bubble radius, the selected TEM images reveal the hydrogen intercalation wave mechanism, which progresses through three distinct stages.
In Stage 1, the $\beta$-$\mathrm{PdH}_x$ nucleates and remains localized at one corner of the Pd nanocube. Once $\smallsim$25\% of the nanocube is transformed into the $\beta$-$\mathrm{PdH}_x$, the phase propagation temporarily stagnates. The electron dose rate was increased to supply more hydrogen and thus promote further hydrogen absorption, as indicated by the larger bubble radius. After this stagnation, the phase transformation enters Stage 2, where the $\alpha$/$\beta$-$\mathrm{PdH}_x$ phase boundary spans the nanocube. This phase boundary forms as the corner-localized $\beta$-$\mathrm{PdH}_x$ expands along the [010] direction. After $\smallsim$50\% of the nanocube is transformed, the $\beta$-$\mathrm{PdH}_x$ propagation slows again until the $\alpha$/$\beta$-$\mathrm{PdH}_x$ interface advances along the [100] direction (Stage 3), completing the phase transformation. This liquid-phase experiment directly visualizes the hydrogen intercalation wave mechanism during hydrogen absorption in the Pd nanocube, which aligns with previously reported absorption mechanisms in Pd nanocubes $\smallsim$20 nm or larger\cite{narayan_direct_2017, sytwu_visualizing_2018}.
\section{Details of elastic Ising model for the $\mathrm{PdH}_x$ system}

\subsection{Parameterization of the Hamiltonian}
The elastic Ising model was used to model the interconversion of the $\alpha$-$\mathrm{PdH}_x$ and $\beta$-$\mathrm{PdH}_x$ phases. This simple model includes harmonic interactions between both nearest and next-nearest neighbors on a simple cubic lattice. The Hamiltonian is given by a sum over lattice positions, $\mathbf{R}$,
\begin{align}
    H = \sum_\mathbf{ R } \Bigg[ \frac{ K_1 }{ 4 } \sum_{ \hat{ \mathbf{ \alpha } }( \mathbf{ R } ) } [ | a\hat{ \mathbf{ \alpha } } + \mathbf{u}_\mathbf{R} - \mathbf{ u }_{ \mathbf{ R } + a\hat{ \mathbf{ \alpha } } } | - l( \sigma_\mathbf{ R }, \sigma_{ \mathbf{ R } + a\hat{ \mathbf{ \alpha } } } ) ]^2 \notag \\
    + \frac{ K_2 }{ 4 } \sum_{ \hat{ \mathbf{ \gamma } }( \mathbf{ R } ) } [ | \sqrt{ 2 }a\hat{ \mathbf{ \gamma } } + \mathbf{u}_\mathbf{R} - \mathbf{ u }_{ \mathbf{ R } + \sqrt{ 2 }a\hat{ \mathbf{ \gamma } } } | - l( \sigma_\mathbf{ R }, \sigma_{ \mathbf{ R } + \sqrt{ 2 }a\hat{ \mathbf{ \gamma } } } ) ]^2  \Bigg]
\end{align}

The set of unit vectors $\{\hat{ \mathbf{ \alpha } }( \mathbf{ R } ) \}$ points from site $\mathbf{R}$ to the nearest neighbors of $\mathbf{R}$ on the simple cubic lattice. For a bulk lattice site, there are six vectors, while generally there are fewer for sites at the edge of the cube. The set of unit vectors $\{\mathbf{\hat{\gamma}}( \mathbf{R} )\}$ point to the next-nearest neighbors of site $\mathbf{R}$, totaling 12 for a bulk lattice site. The variables $\{ \mathbf{u}_\mathbf{R} \}$ represent displacements of site $\mathbf{R}$ from its equilibrium position. When site $\mathbf{R}$ is viewed as a unit cell of the Pd nanocube, $\{ \mathbf{u}_\mathbf{R} \}$ represents a compression or expansion of the cell. The variables $\{ \sigma_\mathbf{R} \}$ are the spins variables and take values $\pm 1$, with $+1$ corresponding to a unit cell that is locally $\beta$-$\mathrm{PdH}_x$ and $-1$ corresponding to a unit cell that is locally $\alpha$-$\mathrm{PdH}_x$. $l( \sigma_\mathbf{R},\sigma_{ \mathbf{R}^\prime } )$ is the preferred distance between two sites, which, due to the expansion of the lattice in the $\beta$ phase, depends on the values of the spin variables. We take $l( \sigma_\mathbf{R},\sigma_{ \mathbf{R}^\prime } )= a_0 + \Delta/2+\Delta/4 (\sigma_\mathbf{R} + \sigma_{ \mathbf{R}^\prime } )$, where $a_0$ is the equilibrium lattice constant of bulk Pd and $\Delta$ is approximately 3.5\% of this value (\textbf{Supplementary Table 1}).

The values for the spring constants, $K_1$ and $K_2$, were chosen such that the model could reproduce the response of bulk Pd to a longitudinal stretch and a pure shear. To do so, we considered a 2 × 2 × 2 cube of sites in the model. The corners of our cube are connected by springs with stiffnesses given by $K_1$ and $K_2$ (\textbf{Supplementary Fig.\ 8}). We assume our system is tiled by such cubes, and we account for the fact that the springs of one cube are shared by the neighboring cubes. This means that the springs with stiffness $K_1$ have a weight of one-fourth, and that the springs of stiffness $K_2$ have a weight of one-half.

If all sites are the same spin, then a configuration with no strain will make all side lengths of the cube equal to $l$, the lattice constant for that phase. We apply a stretch and a shear strain to the cube and calculate the energetic cost according to \textbf{Supplementary Equation 1}. This gives us $\Delta E_1 = (K_1+2K_2)\Delta y^2/2$ and $\Delta E_2 = K_2 \Delta x^2/2$, while elastic theory gives $\Delta E_1 / l^3 =c_{11} (\Delta y/l)^2 / 2$ and $\Delta E_2 / l^3 = c_{44} (\Delta x/l)^2 / 2$. Therefore, a consistent choice for the spring constants are,  $K_1 = l(c_{11} - 2c_{44} )$, and $K_2 = lc_{44}$. Hsu and Leisure\cite{hsu_elastic_1979} have calculated the low-temperature limits of the elastic moduli of $\mathrm{PdH}_x$ in both the $\alpha$ and $\beta$ phases. Since we have created our mapping considering only the energetic cost rather than the free energy, we take the 0 K extrapolations for $c_{11}$ and $c_{44}$. Since we wish to model the regime of coexistence, we take averages of the $\alpha$ and $\beta$ phase values for both the value of $l$ and the elastic moduli. This procedure provides the parameters listed in \textbf{Supplementary Table 1}.

\captionsetup[table]{name=Supplementary Table}
\captionsetup[table]{labelfont=bf}

\begin{table}[htbp]
\centering
\begin{tabular}{ | c | c |}
 \hline
 \textbf{Parameter} & \textbf{Value}  \\ 
 \hline
 $\Delta$ & 0.1387 \AA \\  
 \hline
 $a_0$ & 0.3890 \AA \\  
 \hline
 $K_1$ & 2.2303 eV/$\text{\AA}^2$ \\
 \hline
 $K_2$ & 1.7307 eV/$\text{\AA}^2$ \\
 \hline
\end{tabular}

\caption{Parameters used for the Hamiltonian (\textbf{Supplementary Equation 1}) based on experimental data from the reference.\cite{hsu_elastic_1979}}
\label{table:ST1}
\end{table}

\subsection{Small mismatch approximation and effective interaction potential}
By linearizing the absolute value terms in \textbf{Supplementary Equation 1}, one obtains a Hamiltonian in the limit of a small lattice mismatch between the two phases\cite{frechette_consequences_2019} that is quadratic in the displacement variables. We can make the further assumption that vibrations of the lattice relax quickly compared to typical timescales of diffusion of hydrogen between lattice sites. This allows us to analytically integrate out the displacement variables, yielding an effective interaction potential among only the spin variables\cite{frechette_elastic_2021}, of the form,

\begin{align}
    H_{ \mathrm{ eff } } = \frac{ 1 }{ 2 } \sum_{ \mathbf{ R } ,\mathbf{ R }^\prime } \sigma_\mathbf{R} \mathbf{V}_{ \mathbf{ R }, \mathbf{ R }^\prime}\sigma_{ \mathbf{ R }^\prime } 
\end{align}

where spins are coupled by an effective long-range potential. The spin-spin potential, $V_{ \mathbf{ R } ,\mathbf{ R }^\prime }$, is given by a sum of two terms,

\begin{align}
    \mathbf{V}_{ \mathbf{ R } ,\mathbf{ R }^\prime } = \Bigg( \mathbf{S} - \frac{ 1 }{ 4 }\mathbf{C}^T\cdot \mathbf{D}^{-1} \cdot \mathbf{ C } \Bigg )_{ \mathbf{ R }, \mathbf{ R }^\prime } \notag
\end{align}

a term,

\begin{align}
    \mathbf{S}_{ \mathbf{ R },\mathbf{ R }^\prime } = \frac{ K_1 \Delta^2 } { 16 } \sum_{ \hat{ \mathbf{ \alpha } } ( \mathbf{R} ) } ( \delta_{ \textbf{ R }, \mathbf{ R }^\prime } + \delta_{ \mathbf{ R } + a\hat{ \mathbf{ \alpha } }, \mathbf{ R }^\prime } ) 
    + \frac{ K_2 \Delta^2 }{ 8 } \sum_{ \hat{ \mathbf{ \gamma } } ( \mathbf{ R } ) } ( \delta_{\textbf{ R }, \mathbf{ R }^\prime } + \delta_{ \mathbf{ R } + \sqrt{ 2 }a\hat{ \mathbf{ \gamma } }, \mathbf{R}^\prime } ) \notag
\end{align}

and a product term,

\begin{align}
    \mathbf{C}_{ \mathbf{ R },\mathbf{ R }^\prime } = \frac{ K_1 \Delta }{ 2 } \sum_{ \hat{ \mathbf{ \alpha } } ( \mathbf{ R } ) } \hat{ \mathbf{ \alpha } } ( \delta_{\textbf{ R }, \mathbf{ R }^\prime } + \delta_{ \mathbf{ R } + a\hat{ \mathbf{ \alpha } }, \mathbf{R}^\prime } ) 
    + \frac{ \sqrt{ 2 } K_2 \Delta } { 2 } \sum_{ \hat{ \mathbf{ \gamma } } ( \mathbf{ R } ) } \hat{\mathbf{\gamma} }( \delta_{\textbf{ R }, \mathbf{ R }^\prime } + \delta_{ \mathbf{ R } + \sqrt{ 2 }a\hat{ \mathbf{ \gamma } }, \mathbf{ R }^\prime } ) \notag\\
    \mathbf{D}_{ \mathbf{ R },\mathbf{ R }^\prime } = K_1 \sum_{ \hat{ \mathbf{ \alpha } } ( \mathbf{ R } ) } \hat{ \mathbf{ \alpha } }\hat{ \mathbf{ \alpha } } ( \delta_{\textbf{ R }, \mathbf{ R }^\prime } - \delta_{ \mathbf{ R } + a\hat{ \mathbf{ \alpha } }, \mathbf{ R }^\prime } )  
    + K_2 \sum_{\hat{ \mathbf{ \gamma } } ( \mathbf{ R } ) } \hat{ \mathbf{ \gamma } }\hat{ \mathbf{ \gamma } } ( \delta_{\textbf{ R }, \mathbf{ R }^\prime } - \delta_{ \mathbf{ R } + \sqrt{ 2 }a\hat{ \mathbf{ \gamma } }, \mathbf{ R }^\prime } ) \notag
\end{align}

requiring the inversion of the matrix $\mathbf{D}_{ \mathbf{ R },\mathbf{ R }^\prime }$.
\subsection{Kinetic Monte Carlo algorithm}
For all simulations, the standard Gillespie algorithm\cite{gillespie_stochastic_2007} was used to propagate the dynamics. The rate to go from configuration $\mathbf{C}$ to $\mathbf{C}^\prime$ took the form,

\begin{align}
    k_{ \mathbf{C},\mathbf{C}^\prime } = k_0 \exp⁡ \Bigg \{ -\frac{\beta}{2} \bigg( E( \mathbf{C}^\prime ) - E( \mathbf{C} )\bigg) 
    + \frac{\beta\Delta\mu}{2} \bigg( \big( N(\mathbf{C}^\prime ) - N( \mathbf{C} ) \big) \bigg) \Bigg \} 
\end{align}

$E( \mathbf{C} )$ and $N( \mathbf{C} )$ respectively denote the energy and the number of spin up sites in configuration $\mathbf{C}$. The pre-factor, $k_0$, was constant for all processes, and it sets the units of time for the simulation, $k_0^{-1}=\tau_0$. The allowed processes were different for the umbrella sampling and interfacial propagation simulations, which will be described in\textbf{ Supplementary Notes 3.4 and 4 }.

\subsection{Phase diagrams of the model}
\textbf{Supplementary Fig.\ 9} shows four approximate phase diagrams for nanocubes with side lengths of 10, 15, 20, and 30 sites. At each temperature, at least four independent trajectories were initialized in random configurations with average values of $c_\beta = \sum_\mathbf{R}\frac{\sigma_\mathbf{R} + 1}{2N}$ of 0.75 and 0.25. The trajectories were propagated for at least 500 rejection-free kinetic Monte Carlo sweeps, and up to $10^4$ sweeps in the case of the smaller cubes. The allowed moves were spin flips and nearest neighbor exchanges. The average value of $c_\beta$  was computed along each trajectory after discarding 200 sweeps for equilibration. The two markers for each temperature in \textbf{Supplementary Fig.\ 9} are the averages across the independent trajectories beginning near $c_\beta=0.25$ and $c_\beta=0.75$, respectively. The error bars are three standard errors in the mean across the independent trajectories. The dashed lines are quartic fits to the markers, where the fourth-order coefficient is enforced to be negative.

For a system of infinite extent, one can obtain an analytic form for the interaction potential of \textbf{Supplementary Equation 2}, in a Fourier basis\cite{frechette_chemical_2020}. Using a numerical Fourier transform, we can construct a Hamiltonian of the form,

\begin{align}
    H_\mathrm{eff} = \frac{1}{2} \sum_{
    \mathbf{R},\mathbf{R}^\prime \neq \mathbf{R}}\sigma_\mathbf{R} V^\mathrm{bulk}(\mathbf{R} - \mathbf{R}^\prime)\sigma_{\mathbf{R}^\prime}
\end{align}

The simplest mean-field theory for this Hamiltonian yields the self-consistent equation,

\begin{align}
    \tanh⁡(\beta\bar{V}m) = m
\end{align}
\begin{align}
    \bar{V} = -\sum_{\mathbf{R}^\prime \neq \mathbf{R} }V^\mathrm{bulk}(\mathbf{R}-\mathbf{R}^\prime), \hspace{2mm} m=\frac{1}{N}\sum_\mathbf{R}\sigma_\mathbf{R} \notag
\end{align}

Calculating $\bar{V}$ out to 256 lattice sites in all three Cartesian directions, we obtain a bulk, mean-field critical point, $T_c^\mathrm{MF} = 644.7 \mathrm{K}$. The solid line in \textbf{Supplementary Fig.\ 9} is the numerical solution to \textbf{Supplementary Equation 5} with this value of $\bar{V}$. All simulations are run at 300 K.

\section{Computational details for free energy calculations}
 \subsection{Computational details}
Umbrella sampling calculations were carried out using $c_\beta$ as the order parameter on cubes of side length, 15 sites or approximately 6 nm. 415 umbrellas were used with centers between $c_\beta^*=-0.05$ and $c_\beta^*=1.05$. The biasing potential was of the form $U=KN(c_\beta-c_\beta^ * )^2 / 2$. Three spring constants were used in different ranges of $c_\beta^*$, which are summarized in \textbf{Supplementary Table 2.} Trajectories were run with kinetic Monte Carlo (KMC) dynamics, with spin flips allowed at the boundary of the nanocube and exchanges allowed between any nearest-neighbor sites. 200 rejection-free sweeps were discarded for equilibration in each trajectory. The correlation time for the concentration was assumed to be less than five rejection-free sweeps in each trajectory. Trajectories were between $5\times10^3$ and $5\times10^4$ rejection-free sweeps in length. The weighted histogram analysis method (WHAM)\cite{grossfield_implementation_2011} was used to construct the free energy profile with a tolerance of $10^{-3}$. To estimate the statistical error, trajectories were split into five segments of equal length, and separate WHAM calculations were performed on each data set. Three standard errors in the mean of the five independent calculations are smaller than the line widths of \textbf{Figs.\ 4a,b.}
\begin{table}[htbp]
\centering
\begin{tabular}{ | c | c |}
 \hline
 $\mathbf{ c_\beta^{*} }$ \textbf{range} & \textbf{Spring constant (eV)}  \\ 
 \hline
 $-0.05\text{ to }0.09$ & 0.8402 \\  
 \hline
 $0.10\text{ to }0.40$ & 0.6463 \\  
 \hline
 $0.40\text{ to }0.60$ & 0.4524 \\
 \hline
 $0.60\text{ to }0.90$ & 0.6463 \\
 \hline
 $0.90\text{ to }1.05$ & 0.8402 \\
 \hline
\end{tabular}
\caption{Spring constants used for umbrella sampling for umbrellas with centers in different ranges.}
\label{table:ST2}
\end{table}

\section{Construction of density field}
A configuration in the elastic Ising model is represented by a discrete spin-field, where each site has a value of $\pm 1$. However, in the physical system, hydrogen density should vary continuously over space. To produce the density maps of \textbf{Figs.\ 4c-e}, we used the procedure of Willard and Chandler\cite{willard_instantaneous_2010}, convoluting the discrete density field with Gaussian distributions, such that the density for a given site is equal to,

\begin{align}
    \tilde{c}_\beta(\mathbf{R})=\frac{1}{\mathcal{N}}\sum_\mathbf{R^\prime}\sigma_\mathbf{R^\prime}e^{\frac{-|\mathbf{R^\prime} - \mathbf{R}|^2}{2a^2}}
\end{align}

In \textbf{Supplementary Equation 6}, $a$, is the lattice constant of the model. In practice, the sum was cut off at a distance of five unit cells, and $\mathcal{N}$ is the normalization of the discretized Gaussian over this range. For the cube geometries of \textbf{Figs.\ 4c-e}, hard-wall boundaries are used. This means that, in \textbf{Supplementary Equation 6}, $\sigma_\mathbf{R}=0$ if $\mathbf{R}$ is outside the nanocrystal.
\clearpage
\begin{figure}[htbp]
    \includegraphics[width=0.75\textwidth]{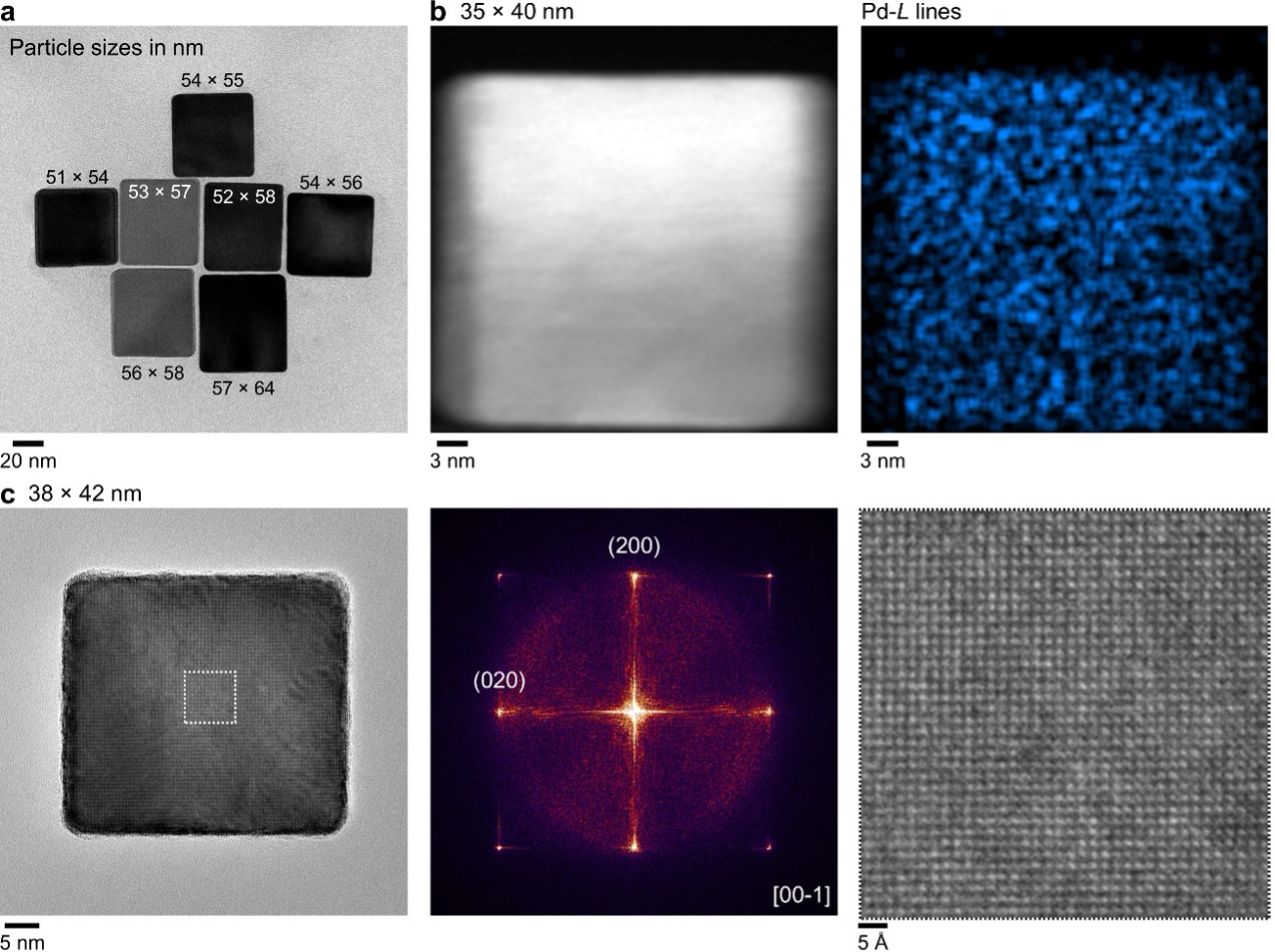}
    \caption*{\textbf{Supplementary Fig.\ 1.} \textbf{S/TEM characterization of as-synthesized Pd nanocubes in the $\smallsim$40 to $\smallsim$60 nm size range.} \textbf{a}, Bright-field TEM image of seven Pd nanocubes, with their sizes labeled in nm. \textbf{b}, STEM image (left) and corresponding Pd-L lines EDS map (right) of a 35 × 40 nm Pd nanocube. \textbf{c}, HRTEM image of a Pd nanocube (38 × 42 nm), with its FFT pattern showing (200) and (020) reflections along the [00-1] zone axis. An enlarged HRTEM image of the selected region (white-dotted area) demonstrates atomic resolution.}
    \label{fig:s1}
\end{figure}
\begin{figure}[htbp]
    \includegraphics[width=0.75\textwidth]{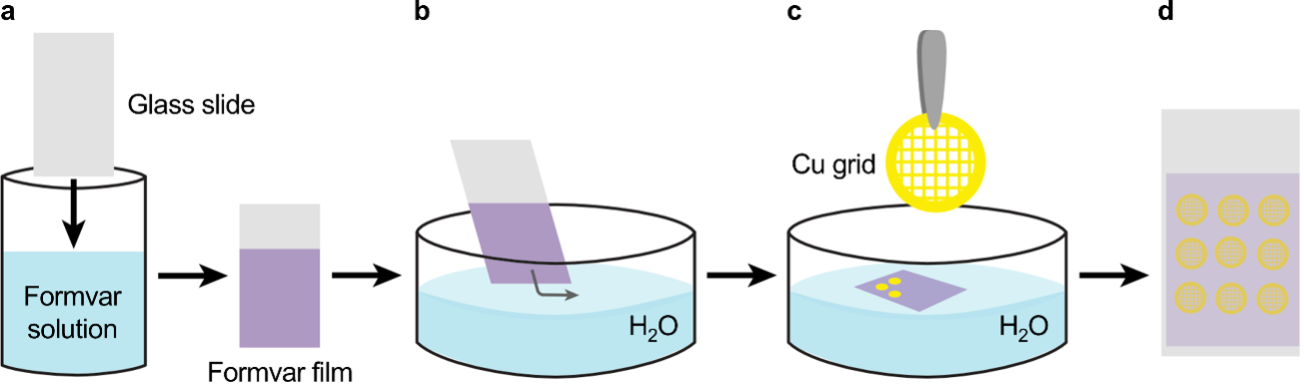}
    \caption*{\textbf{Supplementary Fig.\ 2.}\textbf{ Schematic of the fabrication process for Formvar-coated TEM grids.} \textbf{a}, A glass slide is dipped into a 0.25\% Formvar solution, forming a thin Formvar film on the slide. \textbf{b}, Surface tension is used to detach the Formvar film from the glass slide by submerging it in water. \textbf{c}, The Formvar film floats on the water surface, and TEM grids are carefully placed on top. \textbf{d}, A second glass slide is employed to retrieve the Formvar-coated TEM grids.}
    \label{fig:s2}
\end{figure}
\begin{figure}[htbp]
    \includegraphics[width=0.75\textwidth]{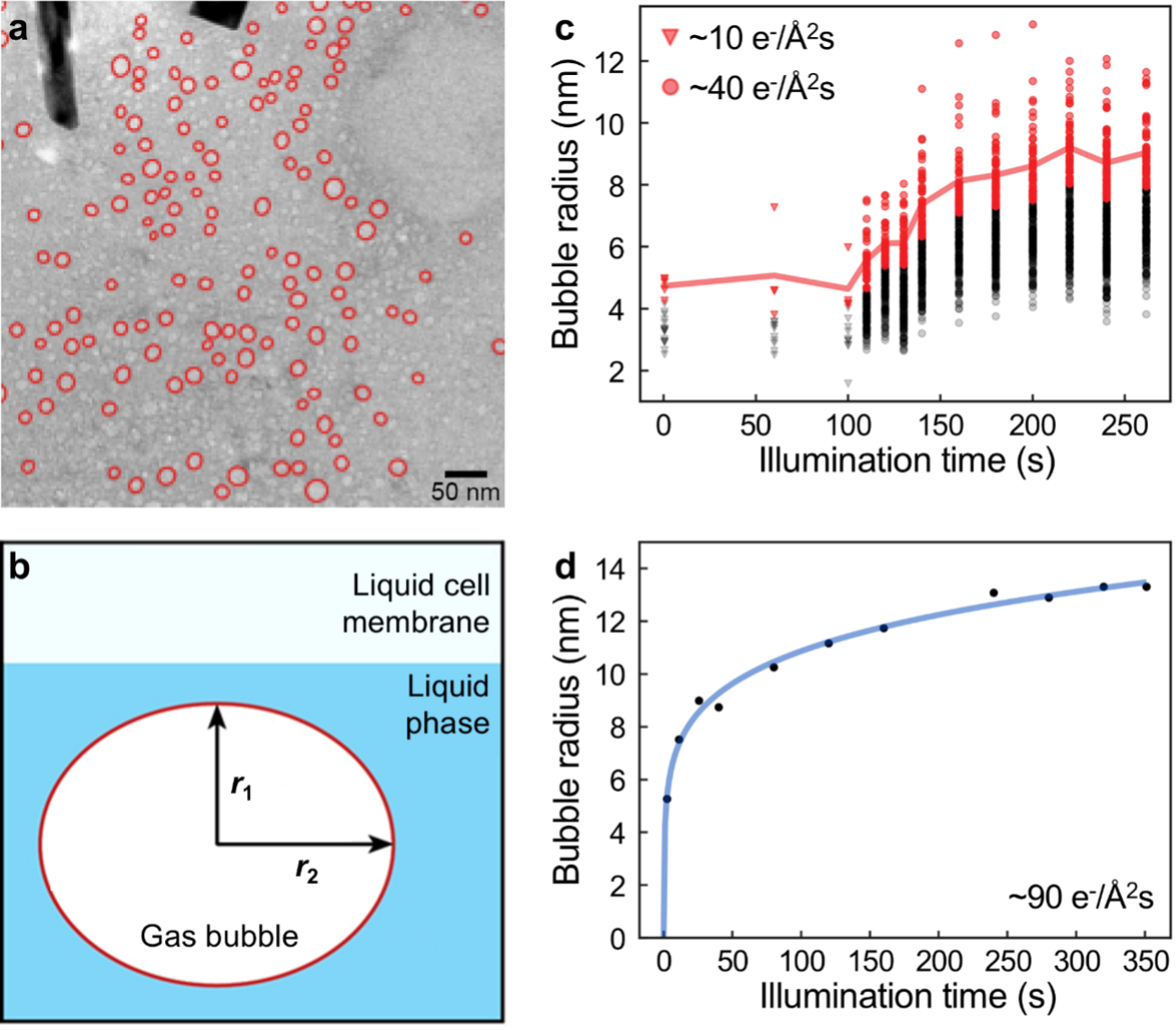}
    \caption*{\textbf{Supplementary Fig.\ 3.} \textbf{Indirect estimation of $\mathrm{H}_2$ concentration inside the liquid cell based on bubble size analysis.} \textbf{a}, A representative TEM image used for bubble size analysis, with the identified bubbles highlighted in red. The TEM image was recorded using a dose rate of $\smallsim$40 e-Å-2s-1. \textbf{b}, Schematic representation of the liquid cell conditions. Ellipses are fitted to the identified bubbles to account for their non-spherical shapes. \textbf{c}, Scatter plot showing the identified bubble radii at different stages of the liquid-phase TEM experiment. The largest 25\% of the radii are highlighted in red and are used to calculate the average radius as a function of reaction time, represented by the red line. The dose rate was increased from $\smallsim$10 to $\smallsim$40 e-Å-2s-1 during the experiment. \textbf{d}, Change in the average bubble radius over time during an experiment conducted at a constant dose rate of $\smallsim$90 e-Å-2s-1. The bubble radii increase rapidly within the first 60 s of illumination, followed by stabilization toward a steady state.}
    \label{fig:s3}
\end{figure}
\begin{figure}[htbp]
    \includegraphics[width=0.50\textwidth]{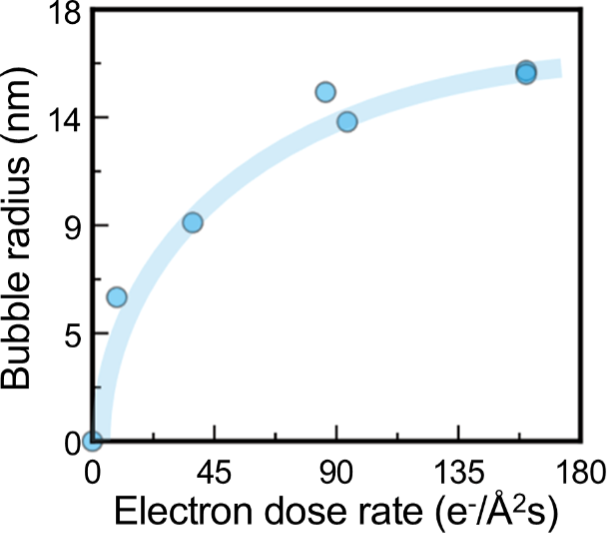}
    \caption*{\textbf{Supplementary Fig.\ 4.} G\textbf{as bubble sizes as a function of electron dose rate.} The steady-state average radius of the gas bubbles varies with the electron dose rate.}
    \label{fig:s4}
\end{figure}
\begin{figure}[htbp]
    \includegraphics[width=0.75\textwidth]{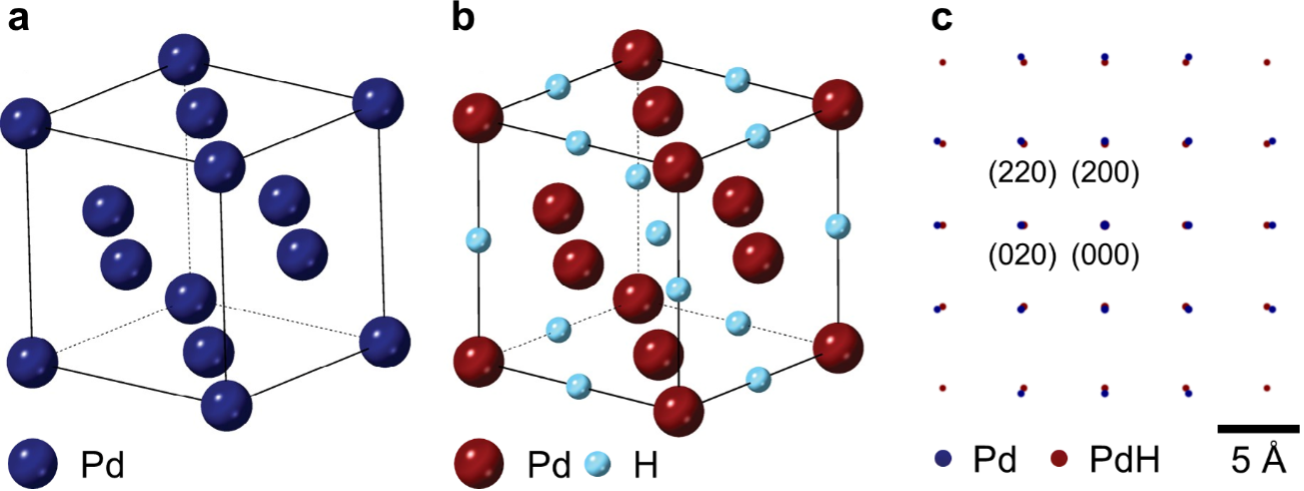}
    \caption*{\textbf{Supplementary Fig.\ 5.} \textbf{Crystal structures of palladium (Pd) and palladium hydride (PdH) with corresponding electron diffraction patterns.} \textbf{a},\textbf{b}, Schematic representations of the crystal lattices of Pd (\textbf{a}) and PdH (\textbf{b}). \textbf{c}, Incorporating hydrogen into the palladium lattice results in lattice expansion, reflected in changes to the lattice parameter\cite{wyckoff_structure_1924, wicke_hydrogen_1978} This expansion is observable in the simulated diffraction patterns, as shown by the difference in radial distances of the reflections for Pd (blue) and PdH (red); the same applies to the FFT patterns. Given that the difference in radial distances between Pd and $\alpha$-$\mathrm{PdH}_x$ is slight, under our in situ liquid-phase experimental conditions, we can consider Pd as equivalent to $\alpha$-$\mathrm{PdH}_x$.}
    \label{fig:s5}
\end{figure}
\begin{figure}[htbp]
    \includegraphics[width=0.75\textwidth]{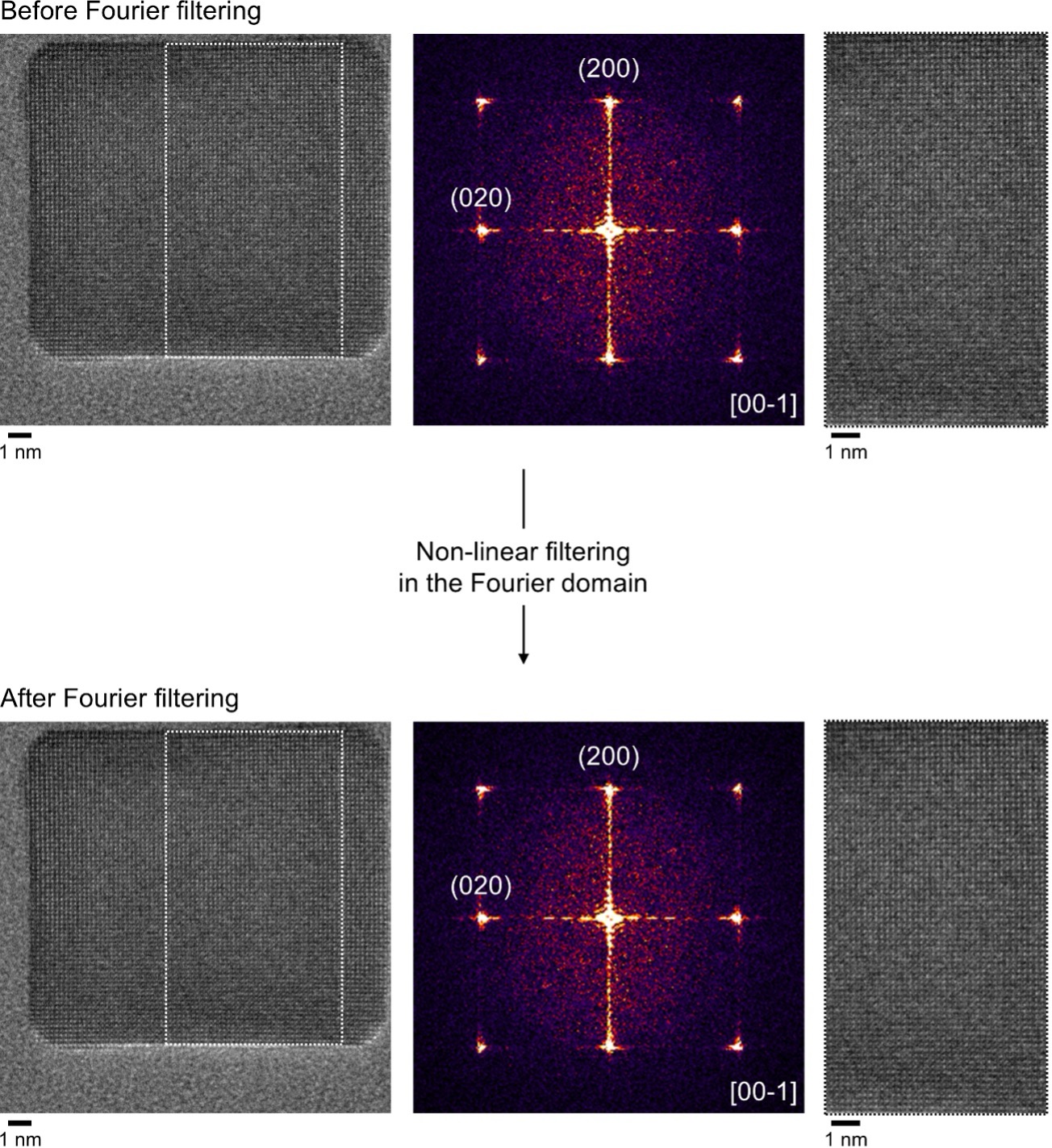}
    \caption*{\textbf{Supplementary Fig.\ 6.} \textbf{Example demonstrating non-linear Fourier filtering of HRTEM images.} HRTEM images, their corresponding FFT patterns, and enlarged HRTEM images of the white-dotted regions are shown before (top row) and after (bottom row) non-linear Fourier filtering. An absolute value shrinkage filter, varying linearly with spatial frequency in the Fourier domain, is applied to attenuate Gaussian noise present in the HRTEM images.}
    \label{fig:s6}
\end{figure}
\begin{figure}[htbp]
    \includegraphics[width=0.75\textwidth]{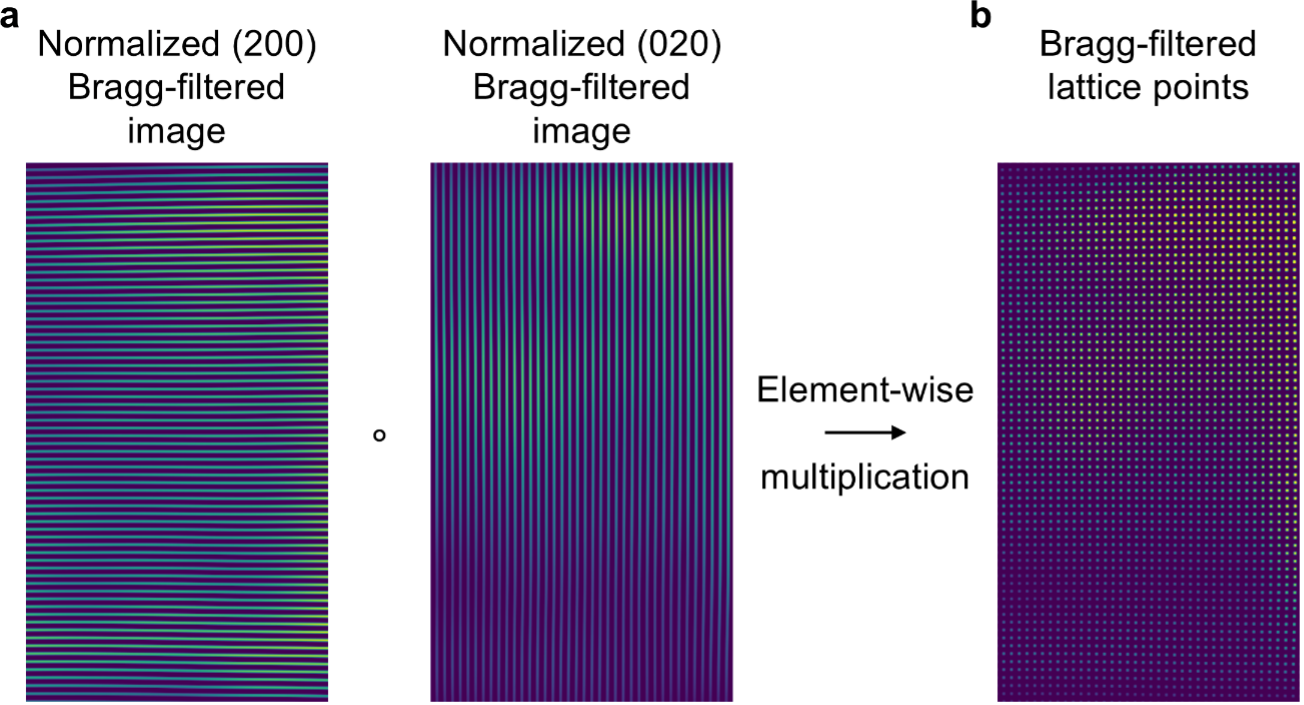}
    \caption*{\textbf{Supplementary Fig.\ 7.} \textbf{Steps for generating Bragg-filtered lattice points.} \textbf{a}, Bragg-filtered images based on the (200) and (020) reflections are clipped to a range from zero to their maximum values (i.e., white contrast). Each clipped image is then normalized between 0 and 1, resulting in normalized (200) and (020) Bragg-filtered images. \textbf{b}, Element-wise multiplication (denoted by `$\circ $' in the figure) of the two normalized Bragg-filtered images yields the Bragg-filtered lattice points.}
    \label{fig:s7}
\end{figure}
\begin{figure}[!htb]
    \includegraphics[width=0.70\textwidth]{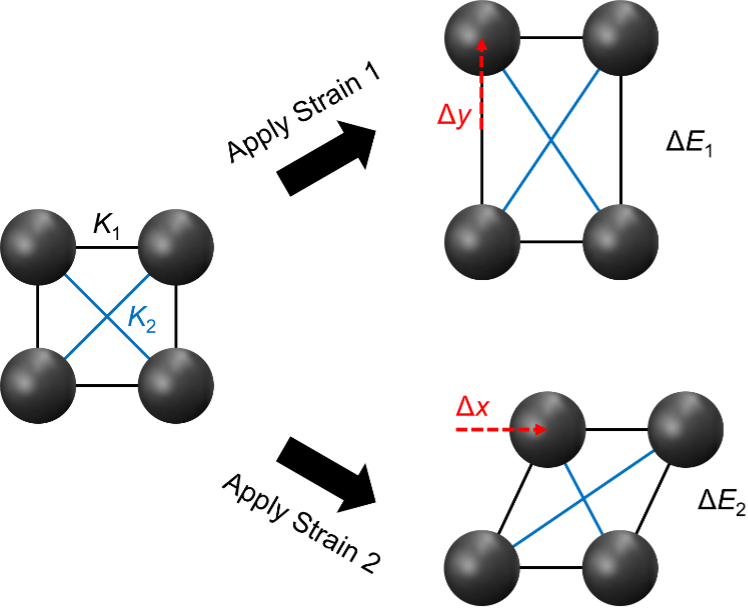}
    \caption*{\textbf{Supplementary Fig.\ 8.} \textbf{Depiction of the two applied strains used to parameterize the model Hamiltonian as described in }\textbf{Supplementary Note 3.1} .}
    \label{fig:s10}
\end{figure}
\begin{figure}[!htb]
    \includegraphics[width=0.70\textwidth]{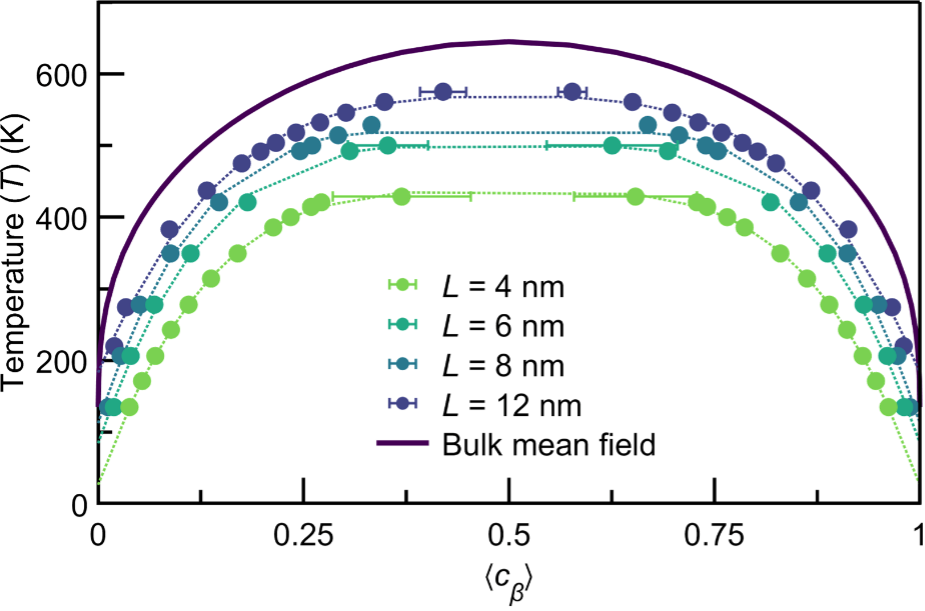}
    \caption*{\textbf{Supplementary Fig.\ 9.} \textbf{Phase diagrams for the elastic Ising model, for different size nanocubes (with side length, $L$).} Markers are averages over at least 4 independent trajectories of lengths of at least 500 rejection-free KMC sweeps. Dashed lines are quartic fits to the markers. The solid line is the numerical solution to an equation of state under a simple mean-field approximation in a bulk system.}
    \label{fig:s11}
\end{figure}
\end{widetext}
\clearpage
\section*{References}
\bibliography{main.bib}

\end{document}